\documentclass[sigconf, nonacm]{acmart}

\newcommand\vldbdoi{XX.XX/XXX.XX}
\newcommand\vldbpages{XXX-XXX}
\newcommand\vldbvolume{14}
\newcommand\vldbissue{1}
\newcommand\vldbyear{2020}
\newcommand\vldbauthors{\authors}
\newcommand\vldbtitle{\shorttitle} 
\newcommand\vldbavailabilityurl{https://github.com/microsoft/prose-benchmarks/tree/main/ConditionalFormatting}
\newcommand\vldbpagestyle{empty} 

\usepackage{algorithm}
\usepackage[noend]{algorithmic}
\usepackage{xcolor}
\usepackage{microtype}
\usepackage{enumitem}
\usepackage{newfloat}
\usepackage{listings}
\usepackage{subcaption}
\usepackage{adjustbox}

\usepackage{graphicx} 
\usepackage{courier} 
\usepackage{lipsum} 
\usepackage{amsmath}
\usepackage{natbib}

\usepackage{booktabs}
\usepackage{tabularx}
\usepackage{xurl}
\usepackage{multirow}

\usepackage{mathtools}

\usepackage{tikz}
\newcommand*\circled[1]{\textcircled{\raisebox{-0.8pt}{#1}}}

\newcommand{\system}[0]{\textsc{DataVinci}}

\DeclarePairedDelimiter{\ceil}{\lceil}{\rceil}

\newtheorem{example}{Example}

\definecolor{jccolor}{rgb}{0.1,0.7,0.8}
\definecolor{vlcolor}{rgb}{0.9,0.1,0.1}
\definecolor{gcolor}{rgb}{0.7,0.3,0.7}
\definecolor{ccolor}{rgb}{0.3,0.3,0.7}
\definecolor{mrcolor}{RGB}{163,96,50}
\definecolor{mscolor}{RGB}{8, 102, 3}

\newcommand\significant[0]{significant}
\begin{document}

\title{\system{}: Learning Syntactic and Semantic String Repairs}

\author{Mukul Singh}
\affiliation{%
  \institution{Microsoft}
  \city{Delhi}
  \country{India}
}
\email{singhmukul@microsoft.com}

\author{Jos\'e Cambronero}
\affiliation{%
  \institution{Microsoft}
  \city{New Haven}
  \country{USA}
}
\email{jcambronero@microsoft.com}

\author{Sumit Gulwani}
\affiliation{%
  \institution{Microsoft}
  \city{Redmond}
  \country{USA}
}
\email{sumitg@microsoft.com}

\author{Vu Le}
\affiliation{%
  \institution{Microsoft}
  \city{Redmond}
  \country{USA}
}
\email{levu@microsoft.com}

\author{Carina Negreanu}
\affiliation{%
  \institution{Microsoft Research}
  \city{Cambridge}
  \country{UK}
}
\email{cnegreanu@microsoft.com}

\author{Gust Verbruggen}
\affiliation{%
  \institution{Microsoft}
  \city{Keerbergen}
  \country{Belgium}
}
\email{gverbruggen@microsoft.com}

\begin{abstract}
String data is common in real-world datasets: 67.6\% of
values in a sample of 1.8 million real Excel spreadsheets from the web were represented
as text. Systems that successfully 
clean such string data can have a significant impact on real users.
While prior work has explored errors in string data, proposed approaches
have often been limited to error detection or require that the user provide annotations, examples, or constraints to fix the errors. Furthermore, these systems 
have focused independently on syntactic errors or semantic errors in strings,
but ignore that strings often contain both syntactic and semantic substrings.
We introduce \system{}, a fully unsupervised string data error detection
and repair system. \system{} learns regular-expression-based
patterns that cover a majority of values
in a column and reports
values that do not
satisfy such patterns as data errors. \system{} can automatically derive
edits to the data error based on the
majority patterns and constraints learned over other columns without the need for further user interaction.
To handle strings with 
both syntactic and semantic substrings,
\system{} uses an LLM to abstract (and re-concretize) portions
of strings that are semantic prior to learning
majority patterns and deriving edits.
Because not all data can result in 
majority patterns, \system{}
leverages execution information from
an existing program (which reads the
target data) to identify and correct
data repairs that would not
otherwise be identified.
\system{} outperforms 7 baselines on both
error detection and repair when evaluated on 4 existing and new benchmarks.
\end{abstract}
\maketitle

\pagestyle{\vldbpagestyle}
\begingroup\small\noindent\raggedright\textbf{PVLDB Reference Format:}\\
\vldbauthors. \vldbtitle. PVLDB, \vldbvolume(\vldbissue): \vldbpages, \vldbyear.\\
\endgroup
\begingroup
\renewcommand\thefootnote{}\footnote{\noindent
}\addtocounter{footnote}{-1}\endgroup


\begin{figure*}[t]
\centering
\includegraphics[width=0.9\textwidth]{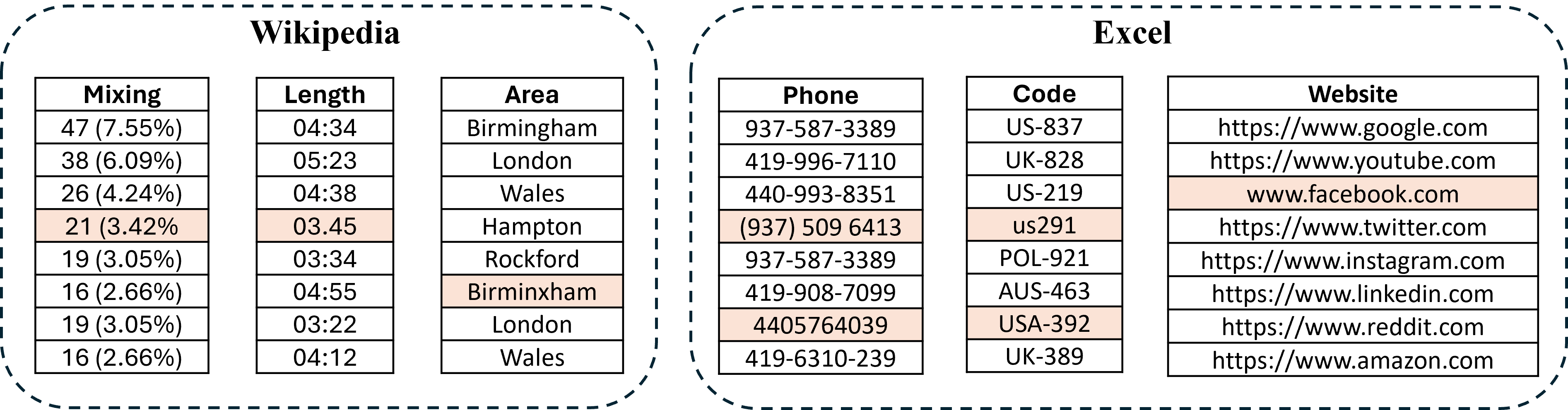}
\caption{Examples of string data errors found in Wikipedia tables and Excel spreadsheets. The data errors in each column are highlighted. There is a mix of both syntactic (like 03.45) and semantic data errors (like Birminxham) found in real tables.
}
\label{fig:error_examples}
\end{figure*}

\section{Introduction}
Errors in tabular data, such as inconsistent 
or corrupted values,
are common and 
can result in 
incorrect computations, invalid conclusions,
and downstream data pipeline failures~\cite{chu2016data}.
Such data errors can stem from a variety of sources including
manual data entry, data integration, and faulty computations.
Prior work~\cite{unidetect} reported that
even in professional settings, such as financial institutions
and consulting firms, up to 24\% of spreadsheets can
have mistakes, including data errors. Figure~\ref{fig:error_examples} shows some examples of errors found in online Wikipedia tables and public Excel spreadsheets.

In a sample of 1.8 million Excel spreadsheets from the web,
we found that 67.6\% of values are represented
as text (compared to numeric or datetime values), providing a substantial opportunity for string repair systems to help real users. 
While prior work has introduced approaches relevant
to string data, these typically have focused primarily
on detecting errors but not
repairing them~\cite{unidetect, autodetect, pfd, holodetect},
required users to provide (partial) annotations
or constraints to drive a semi-supervised detection/repair
procedure~\cite{raman2001potter, holoclean, raha, origraha},
or have relied on a
limited rule-learning approach~\cite{wmrr}.

Furthermore, effective string cleaning must support
errors in columns where values contain
both syntactic
and semantic substrings.
For example, given a column with three values \texttt{[(NY, (Boston), (Miami)]} where the pattern is both semantic (city names) and
syntactic (parenthesized values), a repair system should report the first
entry as a data error and suggest \texttt{(New York)}
as the repaired value.
Unfortunately, prior rule-based
and external-knowledge-based systems can only tackle
the syntactic issue or the semantic issue, respectively,
but not the combination of these.

We introduce \system{}, a fully automated 
error detection
and resolution system for string columns in 
tabular data.
\system{} is designed to handle
qualitative string errors~\cite{origraha} in a column,
such as missing string values, inconsistent
formats, or misspellings in strings.
Critically, in contrast to prior
work~\cite{origraha, autodetect, unidetect},
\system{} does not only detect data errors in strings
but also suggests repairs.
\system{} is the first system,
to our knowledge, to detect and repair errors in
strings that consist of both syntactic
and semantic substrings.
Prior work has been tailored to either category separately.
Finally, \system{} can perform detection and repair
in a fully unsupervised manner, without
requiring user inputs such as providing
constraints~\cite{holodetect}, examples~\cite{origraha},
or annotations~\cite{raman2001potter}. 

\paragraph{Error Detection} 
To carry out fully unsupervised error detection on a string column,
\system{} exploits the regularity in string data, and 
reports as data errors those values that do not satisfy patterns associated with a (configurable) large fraction of the column's values.
In contrast to existing pattern-based work~\cite{wmrr, pfd, autodetect}, \system{} uses an LLM to identify
and mask semantic substrings, allowing the regular-expression-based
pattern learner to capture strings with both syntactic
and semantic substrings.

\paragraph{Error Repair}
While prior string repair systems require that the
user provide examples~\cite{raha}, annotations~\cite{raman2001potter}, or constraints~\cite{holoclean}, \system{}
can suggest data repairs without any additional input
by comparing the data error to the regular-expression-based
patterns (i.e. \significant{} patterns) that are associated with a large fraction
of values.
Specifically, \system{}
generates candidate repairs by
deriving a minimal set of edits to an erroneous
value that lead to satisfying a \significant{} pattern. Because these edits may contain
elements such as character classes that
need to be concretized,
\system{}
learns relationships between non-error values and \significant{} regular expressions
and uses these 
as constraints when
predicting the concrete
values. After these edits are applied any
semantic mask values remaining are concretized
by replacing them with concrete LLM-predicted substrings. \system{} sorts the
final set of candidate repairs based on a heuristic ranker.

\paragraph{Execution-Guided Repair}
By flagging values that do not
satisfy \significant{} patterns as data errors,
\system{} can mitigate
false positives. However, 
real data may not always
display such \significant{}
patterns. For example,
consider a column named {col1}
with values \texttt{[c-1, c-2, c3, c4]}.
A pattern learner would
identify two patterns, 
one to cover the first
two entries and
a second to cover the last two entries.
Without further information, it is not possible to identify 
any of these entries as
a data error under a majority assumption.
However, executing the user-written spreadsheet formula \textsf{=SEARCH(``-'', [@col1])}), which 
searches for the \textsf{``-''} in a string, on the first two values will yield a valid result, but on the last two values will result in an exception.
This execution provides a strong signal that the last two values are data errors with respect to this formula.
\system{} can exploit this execution information to provide suggested repairs.
First, \system{} flags the last two value as
data errors given their exceptional execution. 
Second, \system{} learns a regular expression
exclusively over the two successful input values.
Third, \system{} applies the pattern-based 
repair procedure previously  described to the failing inputs, producing the expected repairs \texttt{c-3} and \texttt{c-4} by inserting the missing ``-'' characters.

We evaluate \system{} on an existing
benchmark (Wikipedia web tables~\cite{autodetect}) and three new benchmarks (a collection of real-world
Excel tables,
synthetically corrupted Excel tables,
and Excel
formulas and their input data).
We compare our performance to 
7 baselines, which include 5 existing
error detection/repair systems,
one LLM-based baseline, and one small transformer-based model.
In addition, we augment detection-only systems
with a GPT-3.5 error repair module.
Jointly these
cover a variety of existing approaches from
the data management community or
reasonable alternatives.

We find that \system{}'s fully unsupervised
pattern-based
detection and repair approach
leads to higher precision detection (+1.6 to +8.7
points than next best), higher recall
detection (+3.6 than next best),
and higher precision repairs (+1.6 to +12.2 points than next best).
We find that removing the LLM-based
abstraction and concretization approach that enables repairing mixed syntactic and semantic strings reduces repair precision and recall (-3.8 and -6 points, respectively).
Finally, we show that applying \system{}'s execution-guided repair 
to our benchmark of Excel formulas
raises formula-level
execution success rates to 
 54\% (single-column inputs) and 47.8\% (multi-column inputs)
compared to 43.2\% and 35.7\% for \system{} without
execution-guided repair, and substantially outperforming
the next best non-\system{} baseline.

To summarize, our contributions are:

\begin{itemize}
    \item We develop a pattern-based approach to 
    detect \emph{and} repair errors in strings with
    both syntactic and semantic substrings.
    \item We implement our approach in 
    \system{}, which can detect data errors and suggest
    repairs in a fully unsupervised
    fashion. We also introduce execution-guided repair
    for data cleaning, where \system{} executes
    an existing program that depends on the target column, 
    uses the resulting execution
    success information to learn patterns,
    and reduces the associated program failures by applying
    suggested repairs.
    
    \item We carry out an extensive evaluation of \system{} on multiple datasets from different domains and against multiple existing systems, which shows that \system{} can outperform existing string error detection and repair approaches.

    \item We release the URLs and data preparation scripts to produce
    our three novel Excel-based benchmarks.\footnote{Due to compliance we are not able to release the post-processed data, so we release scripts replicate our benchmarks.}

\end{itemize}

\begin{figure*}[t]

\centering
\includegraphics[width=0.95\textwidth]{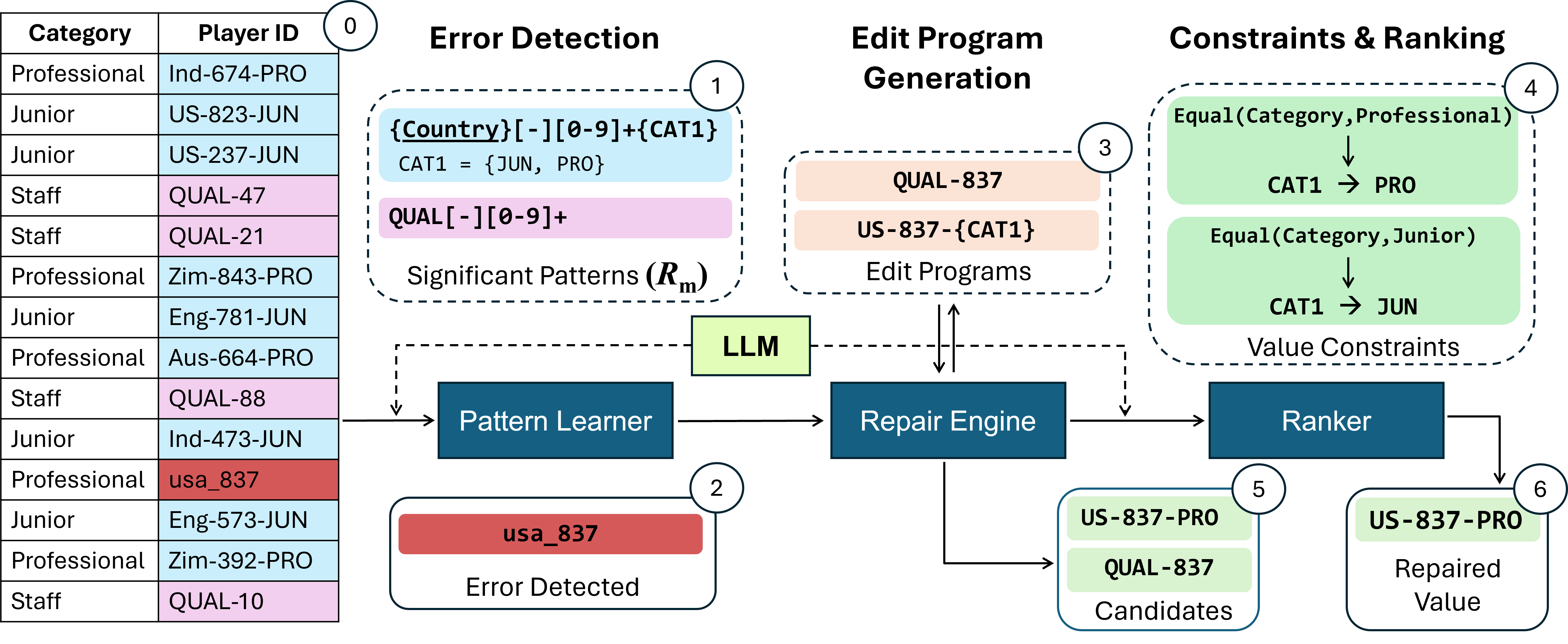}
\caption{\system{} illustrated with an example from our corpus. \textcircled{0} denotes the input table. \textcircled{1} \system{} learns  \significant{} patterns, which account for both syntactic and semantic substrings (using the LLM), for the \emph{Player ID} column. The values satisfying each pattern are shown in the pattern's color. Underlined pattern elements (e.g. Country) correspond to LLM-abstracted semantic substrings. \textcircled{2} \significant{} patterns are used to detect outliers. \emph{usa\_837} is detected as an error as it does not satisfy a \significant{} pattern. \textcircled{3} edit programs for the outlier are learned by deriving edits to the string that will satisfy a \significant{} pattern. \textcircled{4} constraints are generated to concretize 
abstraction actions in edit programs. Here, the constraint between\texttt{\{CAT1\}} and \emph{Category} column is learned. 
Applying these value constraints produces the candidate repairs \textcircled{5}. The heuristic ranker sorts candidates and \textcircled{6} the top ranked candidate is suggested to the user.
}
\label{fig:arch}
\end{figure*}

\section{Problem Statement}
We take inspiration
from the data error detection
and repair formulation
presented in prior work~\cite{autodetect,holoclean}.
First, we introduce our target
domain of string data errors.

\begin{definition}[String Data Error]
Let $T$ be a table,
consisting of a collection of
$m$ columns $\{c_1, \dots, c_m\}$,
each with $n$ values.
Let $c_i$ be a string column consisting
of observed values $\{v_1, \dots, v_n\}$, where $v_j$ is a string value.
For each $v_j$, let $v_j^{*}$ be the 
latent clean value.
If $v_j \neq v_j^{*}$, we say $v_j$ is a string data error.
\end{definition}

The goal of a string repair system is
not only to identify such string errors,
but also provide repaired values.

\begin{definition}[String Error Repair]
Let $v$ be a string value with a data error.
Let $\hat{v}$ be a string value produced by
a data repair system, given $v$ and the table
$T$. We say
$\hat{v}$ is a candidate string error repair. If
$\hat{v} = v^{*}$, we say it is a successful
string error repair.
\end{definition}

Because the space of possible errors
and repairs described previously can be
infinite\footnote{consider you can always extend the string with new characters}, we focus our problem
on \emph{pattern-based regular error detection and repair}
in string values. In Section~\ref{sec:semantic}
we show that rather than limit our scope, simple adaptations to this framing
allows us to perform pattern-based semantic repairs.

\begin{definition}[Regular string column and regular errors]\label{def:regular-errors}
Let $\mathcal{L}_i$ be the (pre-defined) \emph{latent} regular language that characterizes 
\emph{latent} values in column $c_i$, such that
$\forall\ v_j \in c_i: v_j^{*} \in \mathcal{L}_i$. 
We say a value $v_j$ is a regular error,
if $v_j \not\in \mathcal{L}_i$.
\end{definition}

\begin{definition}[Regular repair]
Let $v$ be a regular string error.
We say $\hat{v}$ is a candidate regular string repair, if 
$\hat{v} \in \mathcal{L}_i$. If $\hat{v} = v^{*}$,
we say it is a successful regular string repair.
\end{definition}

The goal of our approach is to 
automatically learn a representation
for $\mathcal{L}_i$, and use this representation
along with our table $T$ to 
produce repair candidates for every $v \not\in \mathcal{L}_i$. 
Note that a direct consequence of Definition~\ref{def:regular-errors}
is that a string data error that is in the latent
regular language, but is not the right value, will not
be detectable in this setting. This is reflected
in \system{}'s design and we will show that despite
this restriction, our approach achieves high performance
on real benchmark tasks. We describe \system{}'s components in greater
detail in the following section.

\section{\system{}}

Figure~\ref{fig:arch} shows a schematic overview of \system{}. 
We start
with a table that contains a string column to clean (Player ID) \circled{0}.
\system{} performs error detection
by learning a set of \significant{} patterns \circled{1}
using an off-the-shelf pattern
learner. Because patterns
may need to represent both syntactic
and semantic substrings, we use an 
LLM to replace semantic substrings
with a mask.
Values that do not
satisfy any \significant{} pattern
are flagged as data errors \circled{2}
and are provided
to the repair engine. The
repair engine produces edit programs by deriving the edits necessary for a data error to satisfy a \significant{} pattern. These edit programs can
have abstract edit actions, such
as deciding what character in a class
or substring in a disjunction (e.g., CAT1) to choose
\circled{3}. We resolve these
choices by learning value constraints
over non-error data \circled{4} and applying
these to the abstract edit actions to 
produce full repair candidates \circled{5}.
Because different \significant{} patterns
may produce different repair suggestions,
we employ a heuristic ranker to return
the top suggestion \circled{6}. 
The following sections describe these steps in detail.

\subsection{Detecting Patterns and Errors}

\system{} uses a set of patterns to describe the regular language that a column represents.
Values that do not match any patterns, and thus not accepted by the language, are marked as errors.
Each pattern is described by a regular expression over all characters encountered in our dataset.
As is standard in regular expressions, we use the following character classes for simplicity of notation: digits, cased and uncased letters, alphanumeric, spaces, alphanumeric with spaces, and the common
recurring character class of \texttt{[0, 1]}.

Given a column $c$, \system{} uses FlashProfile~\cite{padhi2018flashprofile} to learn up to $k$ patterns $R = \{r_1, \dots, r_{k}\}$ such that all values $v$ in $c$ are in the language jointly defined by these patterns $\mathcal{L}_R = \bigcup \mathcal{L}_{r_k}$.
FlashProfile supports disjunction (e.g., \textsf{(cat|dog)} matches ``cat'' and ``dog'') and quantification over groups (e.g., \textsf{([a-z].)+} matches one or more repetitions of a letter and a period).
FlashProfile balances the number of individual patterns with the generality (number of cells covered) of each pattern. We use the default parameters.

From these patterns, \system{} then selects the subset of patterns $R_m \subseteq R$ that individually cover at least a fraction $\delta$ of the values. We refer to these as \emph{\significant{}} patterns.
The union of these \significant{} patterns defines the language $\mathcal{L}_{R_m} = \bigcup_{k \in m} \mathcal{L}_{r_k}$.
\system{} reports any value $v \not\in \mathcal{L}_{R_m}$ as a data error.
We can change the confidence required for \system{} to report a value as an error by changing the threshold $\delta$.

\begin{figure}
    \centering
    \includegraphics[width=0.9\columnwidth]{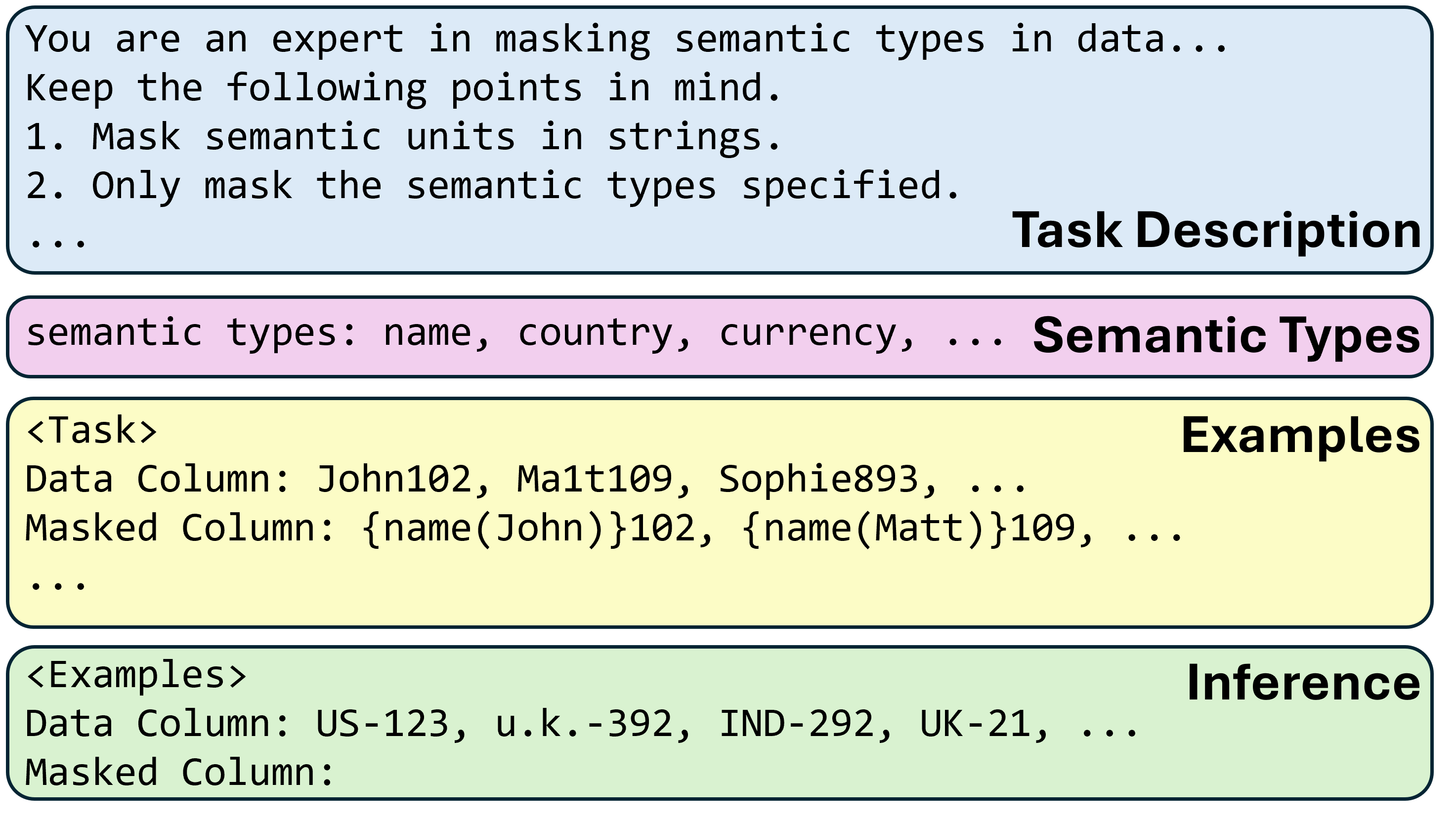}
    \caption{
        \system{}'s semantic abstraction prompt structure. The different colors show the components of the prompt.
    }
    \label{fig:prompt-structure}
\end{figure}

\subsection{Semantic Abstractions} \label{sec:semantic}

To allow \system{}'s repairs to perform
both syntactic and semantic changes, like \texttt{usa\_837} $\rightarrow$ \texttt{US-837} in Figure~\ref{fig:arch} (where the substrings \texttt{usa} and \texttt{US} denote a country),
using a syntactic repair engine, we perform a \emph{semantic abstraction} on each value.
In such an abstraction, substrings that denote some named concept $x$, like a city or a color, are replaced with a mask token $m_x$.

\begin{example}
Consider a column with values [red 1, dark green 2, blue phone 3].
A pattern that matches this column is \textsf{([a-z] )+[0-9]} and we cannot identify that ``phone '' should be removed.
With knowledge about colors, the semantic abstraction of this column is [$m_c$ 1, $m_c$ 2, $m_c$ phone 3].
The \significant{} pattern becomes \textsf{$m_c$ [0-9]} and ``$m_c$ phone 3'' is identified as an error.
\end{example}

One way of obtaining the semantic abstraction of a string is to maintain a dictionary of concepts.
This has three main drawbacks: some concepts are hard to exhaustively enumerate (like colors), spelling mistakes cannot be repaired without additional fuzzy matching (like ``bleu'' instead of ``blue'') and semantic concepts can be contextual (``red'' can also refer to a movie).

We propose to leverage a large language model (LLM) to obtain the semantic abstraction. While abstracting,
we also allow the LLM to 
provide suggestions for
replacement strings
that can be used to 
replace the mask in the
final string.
This allows the LLM to repair spelling mistakes, for example, in Figure~\ref{fig:arch}, it correctly repairs the masked value \texttt{usa} to \texttt{US}.

To capture the context of semantic concepts, we prompt the model with a whole column at once.
Long columns are processed in batches based on the maximum prompt length (4k tokens for GPT-3.5).
Due to the repetitive nature of this task, we found that long prompts did not deteriorate the quality of
generations.

It is important to mask
values at the right level
of granularity.
For example, values from a column [\text{Q4-2002}, \text{Q3-2002}, \text{Q32001}] are masked entirely as \emph{Quarter}
if given to the LLM without
further restrictions. This masking would prevent the last value from being repaired (Q32001 $\rightarrow$ Q3-2001).

To mask values with the right granularity, we only mask a set of predefined semantic categories.
Sherlock~\cite{sherlock}, a prior work on semantic type detection, introduced a method to classify a column as one of 78 popular semantic types, such as
\emph{Name}, \emph{Country} and \emph{Currency}.
We take the 20 most frequently occurring semantic types, which cover 99.2\% of values
with a detected semantic type,
from a sample
25K data columns from our 
Excel data.

Figure~\ref{fig:prompt-structure} summarizes the prompt to perform semantic abstraction  with the LLM.
We use few-shot prompting to show the model both (1) to mask the substring $s$ of semantic type $t$ as $\lbrace t(s) \rbrace$ and (2) that it is allowed to repair the masked values (i.e. $\lbrace t(s') \rbrace$ where $s' \neq s$).
For example, the masked version of ``US-123'' is ``\{country(US)\}-123'' and ``u.k.-392'' becomes ``\{country(UK)\}-392''.
These are then transformed to ``$m_1$-123'' and ``$m_1$-392'' before learning patterns, and $m_1$ is added to the alphabet for our regular expression learner.

\subsection{Repairing Values with Edit Programs}

Given a pattern $r_k \in R_m$ and a value $v \not\in \mathcal{L}_{r_k}$, \system{} repairs $v$ by learning edit programs $e$ such that $e(v) \in r_k$, where we use $e(v)$ to denote applying program $e$ to value $v$.
Let an edit action be a function that optionally deletes a character and optionally emits a given character.
An edit program is then a sequence program over edit actions, which when applied to value $v$, yields one candidate repair.
The edit program is applied to a string by starting from the first character and applying each edit action on the current character and advancing to the next character in the string.
An overview of edit actions is shown in Table~\ref{tab:edit_actions}.

\begin{table}[]
    \centering
    \caption{Edit actions over characters.}
    \label{tab:edit_actions}
    \begin{tabular}{llccc}
        \toprule
        Action                   & Shorthand       & Delete       & Emits & Cost \\ \midrule
        \textsf{match()}         & \textsf{M}      &              &       & 0    \\
        \textsf{insert($c$)}     & \textsf{I($c$)} &              & $c$   & 1    \\
        \textsf{delete()}        & \textsf{D}      & $\checkmark$ &       & 1    \\
        \textsf{substitute($c$)} & \textsf{S($c$)} & $\checkmark$ & $c$   & 1    \\
        \bottomrule
    \end{tabular}
\end{table}

\begin{example}
    Consider an edit program [M, S(2), I(.)] consisting of three steps.
    Highlighting the current character being looked at with an underscore, the string AAA3 is edited as follows.
    $$\text{\underline{A}AA3} \xrightarrow{M} \text{A\underline{A}A3} \xrightarrow{S(2)} \text{A2\underline{A}A3} \xrightarrow{I(.)} \text{A2.\underline{A}A3}$$    
\end{example}

Let $e$ be an edit program that repairs value $v \not\in \mathcal{L}_r$ with respect to pattern $r$.
$e$ is minimal if there does not exist an edit program $e'(v) \in \mathcal{L}_r$ such that $\textsc{dist}(e'(v), v) < \textsc{dist}(e(v), v)$, where \textsc{dist} is the Levenshtein edit distance~\cite{levenshtein} between strings.

\begin{figure}
    \centering
    \includegraphics[width=0.75\columnwidth]{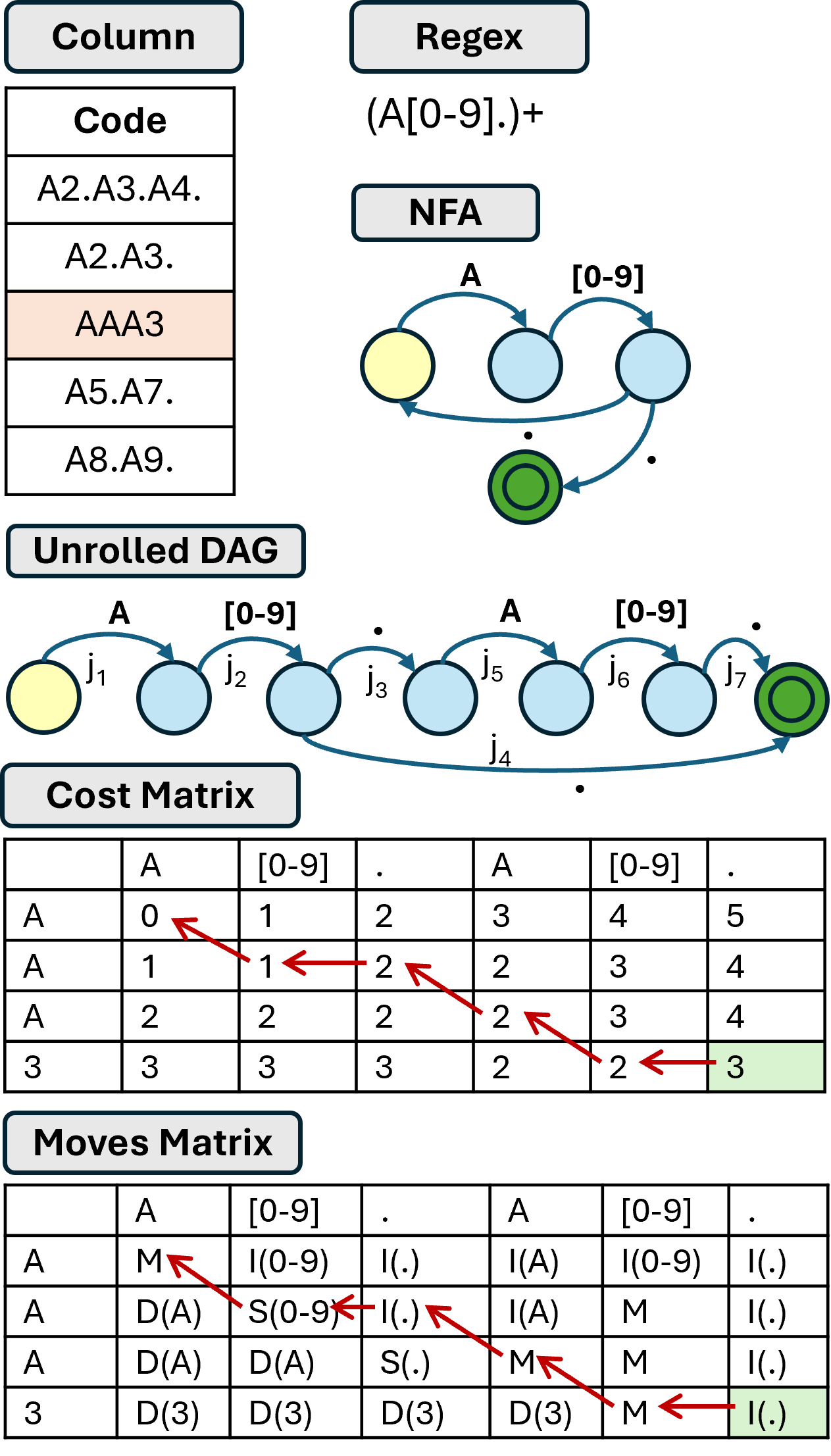}
    \caption{
        \system{}'s repair engine on an example column. The \significant{} regex learned for the column is \emph{(A[0-9].)+)} which is used to identify the outlier (\emph{AAA3}), highlighted in red. The regex is converted to an NFA with the starting and accepting states shown in yellow and green, respectively, and transition symbols over the edges.  The NFA is then unrolled to a DAG for the outlier. We show the computed cost and moves matrix and highlight one optimal repair script path in them with red arrows.
    }
    \label{fig:dp-optimization}
\end{figure}

Given $v \not\in \mathcal{L}_{r_k}$, \system{} learns minimal edit scripts $e(v) \in \mathcal{L}_{r_k}$ using dynamic programming.
The pattern $r_k$ is interpreted as a non-deterministic finite state automaton (NFA) where edges correspond to matching (and consuming) a single character \cite{sipser1996introduction}.
An example of a pattern and its corresponding NFA is shown in Figure~\ref{fig:dp-optimization}.
An erroneous value will end in a non-accepting state
in the NFA where no further transitions can be taken. By changing the characters of the string as we traverse it, edit actions allow us to follow new edges.
These changes come at a cost, however, and these costs are shown in Table~\ref{tab:edit_actions}.
Finding a minimal edit script then corresponds to finding the lowest cost path in the NFA, which is done through dynamic programming.

\begin{example}
    Consider the example in Figure~\ref{fig:dp-optimization}.
    When processing the highlighted data error,
    after transitioning on the \textsf{A} edge, there are no more edges that we can follow.
    For a cost of 1, we can use an edit action \textsf{I(3)} to follow the \textsf{[0-9]} edge.
\end{example}

There are two challenges to finding the lowest cost path: loops due to unbounded quantification and transitioning on character classes and categories. For example, there are ten edit actions \textsf{I(0)} $\cdots$ \textsf{I(9)} that allow following an edge \textsf{[0-9]}.

To handle loops, we approximate the NFA for a given value $v$ with a directed acyclic graph $D_v$ by unrolling loops up to depth $\ceil[\big]{\frac{\textsc{len}(v)}{\textsc{len}(\mathrm{cycle})}}$ with the length of a cycle defined as the number of edges in it.
We support nested cycles and follow the same unrolling procedure recursively for each nested loop.
In practice, 
we found nested loops to 
be rare in our
learned regular expressions.\footnote{less than 1\% of regular expressions learned over 100,000 Excel columns resulted in nested loops}
Figure~\ref{fig:dp-optimization} shows the NFA converted into a DAG by unrolling the loop twice and topologically sorting states.
The loop was unrolled twice as the length of the cycle is 3 and $\ceil[\big]{\frac{4}{3}} = 2$.

To match character classes and disjunctions, we first learn \emph{abstract} edit programs, which have edit actions that emit a character class or a value from a disjunction.
Two examples of abstract edit actions are \textsf{S(0-9)} (see Figure~\ref{fig:dp-optimization}) and \textsf{I(CAT|PRO)} (see Figure~\ref{fig:arch}).
After the minimal abstract edit program is obtained, abstract edit actions are concretized by choosing one of the characters in its character class or one of the strings in the category. Concretization is detailed in Section~\ref{ssec:concretization}---it does not influence how minimal (abstract) edit programs are learned.

Let $\textsc{cost}(i, j)$ be the cost of transitioning on edge $j$ after having consumed $i$ characters in the string and $\textsc{move}(i, j)$ be the edit action required to do so.
In the previous example, we had that $\textsc{cost}(1, j_2) = 1$, as a result of the necessary insertion action.
Since nodes can have multiple incoming edges, we write $p(j)$ to denote incoming edges in the node where edge $j$ starts.
The cost of transitioning on each edge while traversing the string is recursively defined as
\begin{equation}\small\label{dp-eq}
    \textsc{cost}(i, j) = \min
    \begin{cases}
        \min_{j^\prime \in p(j)} \textsc{cost}(i, j^\prime)  + 1 & (\textsc{i})\\
        \min_{j^\prime \in p(j)}\textsc{cost}(i-1, j^\prime) + [s[i] \neq l(j)] & \text{(\textsc{m} or \textsc{s})}\\
        \textsc{cost}(i-1, j) + 1                               & (\textsc{d})\\
    \end{cases}
\end{equation}
with $l(j)$ the label of edge $j$ and $[a \neq b]$ Iverson bracket notation, which evaluates to 1 if $a \neq b$ else 0. 
The associated moves are shown on the right of each case.

\begin{example}
    In Figure~\ref{fig:dp-optimization}, 
    we can arrive at \textsc{cost}(2, $j_2$) (i.e. traversing $j_2$ after consuming two characters) through 3 paths.
    We can move from \textsc{cost}(1, $j_1$)
    to the new state by substituting
    the second A with a digit (S(0-9) with cost 1),
    inserting a digit
    and deleting the second
    A (I(0-9), D with a cost 2),
    or 
    delete the second A and
    insert a digit (D, I(0-9) with a cost
    2). Because substitution had
    the lowest cost (1), 
    \textsc{cost}(2,  $j_2$) = 1 and \textsc{moves}(2, $j_2$) = \textsf{S(0-9)
    }.
\end{example}

The correctness of our DP algorithm can be proven through extension of string edit-distance \cite{stanford-edit}.
The time complexity of the algorithm is $O(m^2n)$ with $m$ the number of edges in the DAG and $n$ the number of characters in the erroneous string $v$.
The space complexity is $O(mn)$ as we only need to store the cost and moves matrices.

\subsection{Concretizing Edit Programs}\label{ssec:concretization}

\system{} learns decision trees to predict concrete values for each character class (which are just disjunctions over a set of characters) and string disjunction in the pattern $r_f$ that induced our edit program.
We refer to these learned rules
as concretization constraints.
By construction, every abstract edit action corresponds to a character class (or string disjunction) in $r_f$. We thus learn a decision tree that uses features from rows where value $v \in \mathcal{L}_{r_f}$ to predict 
the value that allowed transitioning on an edge
in the unrolled DAG that has the
target character class (or string disjunction).

\begin{example}
    Consider value ``A2.A3.'' in row 2 from Figure~\ref{fig:dp-optimization} and the associated 
    unrolled DAG, which has
    two \textsf{[0-9]}
    edges, which match
    2 and 3, respectively.
    Similarity, value ``A5.A7.'' on row 4 matches 5 and 7.
    For the first edge, this yields  two training examples
    $$\text{row 2} \rightarrow 2 \qquad
      \text{row 4} \rightarrow 5\qquad
      $$
    for the decision tree.
\end{example}

To learn decision trees over these training examples, \system{} first extracts boolean features from each row.
We take inspiration from the \textsc{Cornet}~\cite{cornet} system for conditional formatting in tables and generate predicates over a set of templates to use as features.
Table~\ref{tab:predicates} shows all supported predicate templates.
To generate candidate string constants $s$, \system{} considers the set of column values and tokens after splitting (separately) on non-alphanumeric characters, case changes, and switches between contiguous alphabetic and numeric characters. 
For length($v, n$) we consider the top 5 most frequent cell lengths in the column.
In tables with multiple columns, \system{} 
generates these predicated-based features over every
column.

\begin{example}
Consider the first row in Figure~\ref{fig:arch}. 
For the Player ID column and \textsf{TextContain(c, s)}, we generate four constants for $s$. The first is the value itself (\textsf{Ind-674-PRO}).
Splitting the cell obtains tokens \{\textsf{Ind}, \textsf{674}, \textsf{PRO}, \textsf{-}\}. As \textsf{TextContain(Player ID, -)} is true for all cells in the column, this is not considered and dropped. We get four features from the first row: 
\textsf{TextContains(Player ID, Ind)},
\textsf{TextContains(Player ID, 674)},
\textsf{TextContains(Player ID, Pro)},
\textsf{TextContains(Player ID, Ind-674-Pro)}.
\end{example}

\begin{table}[htb]
    \centering
    \caption{Supported predicate templates and their arguments. The $v$ argument denotes the column value. For example, \textsf{equals(col1, "AR")} matches the cells in column \emph{col1} which are equal to \textit{"AR"}.}
    \label{tab:predicates}
    \begin{tabular}{lll}\toprule
      \textsf{equals($v$, $s$)}& \textsf{contains($v$, $s$)}& \textsf{startsWith($v$, $s$)} \\
      \textsf{endsWith($v$, $s$)}& \textsf{length($v$, $n$)}& \textsf{hasDigits($v$)} \\
      \textsf{isNum($v$)}& \textsf{isError($v$)} & \textsf{isFormula($v$)} \\
      \textsf{isLogical($v$)} & \textsf{isNA($v$)} & \textsf{isText($v$)} \\  
    \bottomrule
    \end{tabular}
\end{table}

To learn each decision tree, \system{} samples trees with varying number of split nodes and depth, filters
down to those with an accuracy of at least
$\alpha$ (default 0.8), ranks trees in
ascending order of (nodes, depth), and takes
the first such tree.
This tree can now be applied
to a repair that
has abstract edits to 
predict the concretized
candidate repair.

\subsection{Ranking Repair Candidates}
\label{sec:ranking}

Our repair procedure can produce multiple edit programs (since there may be multiple \significant{} patterns).
To address this challenge, \system{} uses a heuristic candidate ranker.
This heuristic corresponds to a weighted linear combination of edit script properties.
The weights are manually set based on qualitative analysis on a small held-out set of 100 columns sampled from our corpus of Excel spreadsheets.
The four properties are (1) string edit distance between erroneous value and the repaired value, (2) count of alphanumeric edit operations, (3) string edit distance of repaired value to closest value in column, and (4) fraction of column matching the significant pattern used to generate the repair.

\subsection{Execution-Guided Repair}
\system{}'s pattern-based detection relies on 
the assumption that the \significant{} patterns characterize
the data distribution well, and that values that
do not satisfy such patterns are data errors. However,
not all string data will be able to produce a \significant{} patterns nor will \significant{} patterns learned over all values
necessarily highlight errors. To address this challenge,
\system{} can exploit execution information from programs that operate on columns to further refine its error detection and repair suggestions.
This improvement comes from learning a \significant{} pattern set $R_m$ that accounts for different execution outcomes.
We now describe this in detail.

Let $P$ be a program that reads a subset of columns in a table, including our target cleaning column $c$.
We say $P$ is a column-transformation program if it can execute over each row tuple independently
and produces one or more output values for each row---thus generating one or more output columns.

\begin{example}
Consider a table with two
columns 
c1 = [x, y, z] and
c2 = [a, b, c]. 
A program \textsf{concat(c1, c2)},
which produces [xa, yb, zc],
is a column-transformation program, while
\textsf{first(c1)}, which produces
x is not.
\end{example}

\system{} executes the column-transformation program
on $T$ and groups executions into successes and
failures (as signaled by exceptional values, such as nan, or program exceptions).
The non-exception group is then provided
to our regular expression learner and all patterns
learned are treated as \significant{} patterns $R_m$.
All values $v$ in our target column that were inputs to
a failing execution are identified as
data errors, and we apply the repair procedure
previously described. 

\section{Evaluation Setup} \label{sec:eval_setup}
We first describe
the hardware specifications used for carrying out experiments, the benchmarks
we evaluate over and
the baselines to which we compare.\footnote{Data will be released for the camera ready to undergo required compliance checks.}

\subsection{Hardware Specifications}
All experiments were carried out using Python (version 3.8.7) on a machine with an Intel Core i7 processor (base at 1.8 GHz), K80 GPU, 64-bit operating system, and 32 GB RAM.

\begin{table}[tb]
\small
\centering
\caption{Benchmark properties and metrics reported. \# Rows/\# Cols denote the average number of rows/columns in the table.
}
\label{tab:benchmark-tasks}
\begin{adjustbox}{max width=\columnwidth}
\begin{tabular}{llrrr}
\toprule
  Dataset             &  Metrics                &  \# Tables  &  \# Col  & \# Row  \\ \midrule
  Wikipedia Tables    &  Precision, Fire Rate   &  1000       &  5.1     & 27.3    \\
  Excel               &  Precision, Fire Rate   &  200        &  1.6     & 523.4   \\
  Synthetic Errors    &  Precision, Recall, F1  &  1000       & 4.3      & 447.5   \\
  Excel Formulas      &  Execution Success      &  11000      &  1.4     & 216.5   \\ \bottomrule
 \end{tabular}
\end{adjustbox}
\end{table}

\subsection{Benchmarks} \label{sec:benchmarks}

We evaluate on four benchmarks.
We use a benchmark of web tables from prior work~\cite{autodetect} and we collect and release three new Excel-based benchmarks.
We briefly describe the statistics of these benchmarks in Table~\ref{tab:benchmark-tasks}:

\begin{itemize}
    \item \textbf{Wikipedia Tables}: 
    We build on the Wikipedia tables dataset
    released in the original Auto-Detect~\cite{autodetect}
    paper. Following their approach, we take the sample
    of 1000 tables on which they manually annotated
    system predictions and extend this with manual 
    annotations for \system{} and our baselines.
    Like prior work, given this annotation approach, we only report precision~\cite{autodetect}.

    \item \textbf{Excel}: We sampled 200
    tables present in workbooks drawn from a corpus of 1.8 million publicly available Excel workbooks from
the web. 
    Similar to the Wikipedia benchmark, we run
    all available systems on the sampled tables, manually annotate their suggestions,
    and report
    precision.

    \item \textbf{Synthetic Errors}: We 
    sample 1000 Excel tables (disjoint from 
    \textbf{Excel} benchmark) from the same corpus previously described. We then
    synthetically introduce errors
    with the goal of measuring recall.
    To 
    introduce errors, we apply the following
    noise operations: (1) random character insertion, deletion and change, (2) random delimiter insertion deletion and change, (3) random digit swap, (4) random shuffle of characters, (5) random capitalization, (6) random decimal, comma swap in numerics, (7) visually-inspired typos \{$o\rightarrow$0, $l\rightarrow1$, $e\rightarrow3$, $a\rightarrow4$, $t\rightarrow$7, $s\rightarrow5$\}. We 
    randomly 
    corrupt cells
    with 20\% probability.
    For each of the cells to be corrupted, there is a $25\%$ probability of applying 1, 2, 3 or 4 noise operations, sampled
    without replacement from the set of operations described.
    Because it is likely
    there are already real data errors present in 
    the data, systems may detect errors or suggest
    repairs for cells beyond our synthetically corrupted cells. As a result, we
    focus our analysis on recall of
    our synthetic errors, but report
    precision (which will be naturally deflated)
    and F1 for completeness.

    \item \textbf{Excel Formulas}: 
    We create a dataset of the form
    (formula, input columns), where
    formula is an Excel formula
    used to define a column (i.e.
    all rows have the same formula, modulo
    input values), and input columns correspond
    to the input values necessary to execute
    the formula. We restrict ourselves
    to formulas where input values
    and output value are part of the same
    table. The task is to repair
    any data errors in the input columns
    such that the formula evaluates without
    producing any error values. To
    construct this dataset we sampled 15,000
    tables from Excel corpus previously described,
    extracted 11,000 formulas where
    at least 1 cell and less then 25\% 
    of cells result in an error value.
    Of these 11,000 formulas, 7,200
    have a single column input and
    3,800 have multiple column
    inputs (on average 3.4). We use this dataset to evaluate the impact of execution-guidance in string data cleaning.
\end{itemize}

\begin{table}[tb]
\centering
\caption{
System comparison overview. Category denotes the task for which the systems were designed. In the case, of
T5 and GPT-3.5 category reflects our usage in this work.
}
\label{tab:baseline-info}
\begin{tabularx}{\columnwidth}{XX}
\toprule
  System    &  Category   \\ \midrule
  WMRR    &  Detection + Repair   \\
  HoloClean     &  Detection + Repair   \\ \midrule
  Raha    &  Semi-supervised Detection   \\
  Auto-Detect    &  Detection   \\
  Potters-Wheel    &  Interactive Detection+Repair   \\ \midrule
  T5    &  Detection + Repair   \\
  GPT-3.5    &  Detection + Repair   \\ \midrule
  \system{}    &  Detection + Repair   \\ \bottomrule
\end{tabularx}
\end{table}

\begin{table*}[tb]
\centering
\caption{Error detection performance across
datasets. \system{} outperforms in terms of precision
on Wikipedia and Excel, and in terms of recall on our synthetic benchmark.
We report Potter's Wheel and Auto-Detect
results using the annotations released with the Auto-Detect paper.  For synthetic benchmarks the groundtruth is taken as the original
table which can also have inherent data errors which will skew the precision and F1 scoreas explained in Section~\ref{sec:eval_setup}, hence these metrics are reported
with a (*).
}
\label{tab:detection}
\begin{tabularx}{0.9\textwidth}{X c *{2}{c} c *{2}{c} c *{3}{c}}
\toprule
\multicolumn{1}{c}{\multirow{2}{*}{\textbf{System}}} & \multicolumn{2}{c}{\textbf{Wikipedia}} & \multicolumn{2}{c}{\textbf{Excel}} & \multicolumn{3}{c}{\textbf{Synthetic}} \\ 
\cmidrule(lr){2-3} \cmidrule(lr){4-5} \cmidrule(lr){6-8}
& \multicolumn{1}{l}{\textbf{Precision}} & \multicolumn{1}{c}{\textbf{Fire Rate}} & \multicolumn{1}{l}{\textbf{Precision}} & \multicolumn{1}{c}{\textbf{Fire Rate}} & \multicolumn{1}{l}{\textbf{Precision\textsuperscript{*}}} & \multicolumn{1}{c}{\textbf{Recall}} & \multicolumn{1}{c}{\textbf{F1 Score\textsuperscript{*}}} \\ \midrule
WMRR & 70.0 & 2.93\% & 65.8 & 2.76\% & 55.3 & 66.8 & 60.5 \\
HoloClean  & 67.0 & 3.87\% & 65.2 & 2.50\% & 52.1 & 64.1 & 57.5  \\ \cmidrule{1-8}
Raha & 68.9 & 4.03\% & 66.4 & 3.74\% & 59.5 & 68.2 & 63.6 \\ 
Potters-Wheel & 66.2 & -- & -- & -- & -- & -- & -- \\
Auto-Detect & 78.5 & -- & -- & -- & -- & -- & -- \\
\cmidrule{1-8}
T5 & 60.8 & 27.47\% & 53.8 & 19.02\% & 40.5 & 56.3 & 47.1 \\
GPT-3.5 & 73.9 & 10.99\% & 60.4 & 11.71\% & 50.1 & 69.8 & 58.3 \\
\cmidrule{1-8}
\system{} & 80.1 & 16.85\% & 75.1 & 14.39\% & 67.4 & 73.4 & 70.3 \\ 
\bottomrule
\end{tabularx}
\end{table*}

\subsection{Baselines}

We compare against various baselines as summarized in Table~\ref{tab:baseline-info}:
\begin{enumerate}

    \item WMRR~\cite{wmrr}: an unsupervised approach to learn
    weighted data rectifying rules based on 
    functional dependencies. 
    Since the tool is not publicly available, to evaluate against WMRR
    we reimplement it based
    on their paper description. 
    
    \item HoloClean \cite{holoclean}: a popular data repair tool based
    on probabilistic inference, which can repair qualitative and statistical errors. We
    run the code released by the authors on GitHub.
    HoloClean originally requires
    that users provide denial constraints. To evaluate in a fully
    unsupervised setting, comparable to \system{},
    we use a single vacuous denial constraint (specifically, column 1 = column 1).

    \item Raha~\cite{raha}: an ensemble-like
    system that combines multiple error detection
    systems and semi-supervision to train
    an error detection system.
    As Raha requires the user to annotate 
    examples, in our evaluation we take
    the first (top-to-bottom) 5 groundtruth
    errors per column and provide these as examples.
    
    \item Auto-Detect\cite{autodetect}: a co-occurrence-based
    error detection system, which also uses regular-expressions to generalize values. We use the Wikipedia results 
    released with the paper. Unfortunately, we are unable to run the tool on all our benchmarks as it is not available publicly
    and as a result only report performance
    on Wikipedia.
    
    \item Potter's Wheel~\cite{raman2001potter}: a seminal data error
    detection and (semi-supervised) correction system
    based on functional dependencies. 
    We leverage the  Potters Wheel's
    error detection annotations on the Wikipedia dataset released in the original
    Auto-Detect paper.
    Since we do not run Potter's Wheel system and rely on the Wikipedia annotations released, we only
    report Potter's Wheel results on 
    the Wikipedia benchmark.
    
    \item T5\cite{T5}: a popular transformer-based 
    encoder-decoder model pretrained on text. We fine-tune T5  for the task of data repair. Since T5 is a text generation model we encode each column as a stringified list of column values separated by a \texttt{[SEP]} token. The model is trained end to end to 
    generate the repaired column, given the
     potentially noisy column as input. The training data 
     consists
    of 100K dirty samples (generated by the same approach used in our synthetic benchmarks) and we task it with generating the original columns.
    Because we run T5 on a single column at a time, it does not consider other columns while repairing.

    \item GPT-3.5\cite{gpt3}: a state-of-the-art 
    transformer-based decoder-only model. We use the same input structure used to train T5 to include the target column in GPT-3.5's prompt. We use GPT-3.5 in a fewshot setting, providing three static examples of a dirty column and the cleaned output the model needs to generate.
    The static examples are from Excel (disjoint from our benchmarks) and are hand annotated.
    We report results at temperature 0 and top-1 generation.
    After experimenting with multiple temperatures,
    we found temperature 0 works best on average based on precision and F1 score across all benchmarks.
    
\end{enumerate}

To evaluate unsupervised repair when using detection-only systems, such as Raha, Auto-Detect and Potter's Wheel, we add a call to GPT-3.5 where we include the outlier value and its column header along with 10 sample values selected based on spatial proximity (5 rows above and below and 3 columns to the left and to the right with headers).
We ask the model to generate the repaired value. We sample values to fit in the fixed prompt length of 4,000 tokens and make individual repair calls for each outlier detected.

\section{Results and Discussion}
We explore the following research questions:

\begin{itemize}
    \item[RQ1.] Can \system{} accurately detect string errors?
    \item[RQ2.] Can \system{} accurately repair string errors?
    \item[RQ3.] Can \system{} use program execution to improve repairs?
    \item[RQ4.] How do \system{}'s design decisions impact performance? 
\end{itemize}

\begin{table*}[tb]
    \centering
    \caption{
        Error repair performance across datasets. As described in 
        Section~\ref{sec:eval_setup}
        for the Wikipedia and Excel
        benchmarks we report
        repair precision split into
        (1) Certain: repairs that are certain (based on 
        hand annotation) and, (2) Possible: repairs
        that are reasonable but 
        groundtruth cannot be 
        uniquely determined.
        For synthetic benchmarks the groundtruth is taken as the original table which can also have inherent data errors which will skew the precision and F1 score hence, these metrics are reported with a (*).
        Recall is computed as the percentage of cases correctly repaired out of the total errors synthetically introduced.
    }
    \label{tab:repair}
        \begin{tabularx}{0.9\textwidth}{X c *{2}{c} c *{2}{c} c *{3}{c}}
        \toprule
        \multicolumn{1}{c}{\multirow{3}{*}{\textbf{System}}} & \multicolumn{2}{c}{\textbf{Wikipedia}} & \multicolumn{2}{c}{\textbf{Excel}} & \multicolumn{3}{c}{\textbf{Synthetic}} \\ 
        \cmidrule(lr){2-3} \cmidrule(lr){4-5} \cmidrule(lr){6-8}
        & \multicolumn{1}{l}{\textbf{Precision}} & \multicolumn{1}{c}{\textbf{Precision}} & \multicolumn{1}{l}{\textbf{Precision}} & \multicolumn{1}{c}{\textbf{Precision}} & \multicolumn{1}{l}{\multirow{2}{*}{\textbf{Precision\textsuperscript{*}}}} & \multicolumn{1}{c}{\multirow{2}{*}{\textbf{Recall}}} & \multicolumn{1}{c}{\multirow{2}{*}{\textbf{F1 Score\textsuperscript{*}}}} \\
        & \multicolumn{1}{l}{\textbf{(Certain)}} & \multicolumn{1}{c}{\textbf{(Possible)}} & \multicolumn{1}{l}{\textbf{(Certain)}} & \multicolumn{1}{c}{\textbf{(Possible)}} &  &  &  \\ \midrule
        WMRR & 61.1 & 57.8 & 59.2 & 55.6 & 43.2 & 61.1 & 50.6 \\
        HoloClean  & 58.4 & 55.6 & 59.0 & 54.9 & 41.3 & 58.6 & 48.5  \\ 
        \cmidrule{1-8}
        Raha + GPT-3.5 & 58.6 & 54.8 & 56.4 & 53.5 & 45.2 & 62.0 & 52.3 \\ 
        Potter's-Wheel + GPT-3.5 & 56.2 & 52.0 & - & - & - & - & - \\
        Auto-Detect + GPT-3.5 & 66.9 & 63.3 & -- & -- & -- & -- & -- \\
        \cmidrule{1-8}
        T5 & 41.0 & 37.8 & 37.7 & 35.2 & 27.9 & 47.0 & 35.0 \\
        GPT-3.5 & 63.9 & 55.5 & 52.1 & 48.9 & 38.2 & 63.8 & 47.8 \\
        \cmidrule{1-8}
        \system{} & 71.3 & 64.9 & 71.2 & 64.6 & 54.1 & 68.9 & 60.6 \\
        \bottomrule
    \end{tabularx}
\end{table*}

\subsection{Error Detection (RQ1)}

Like prior work~\cite{autodetect,holoclean,raha} we report
precision for detection.
We leverage existing annotations where possible
and otherwise manually annotate systems' predictions. We report precision, recall,
and F1 on our synthetically corrupted
dataset, computed with respect to our
corruptions, as a result
precision/F1 score can be deflated
from preexisting data errors.

In addition to standard metrics
like precision, we also report
each system's \emph{average fire rate}. 
We define this as the average
fraction of cells in a column that
are labeled as data errors.

Table~\ref{tab:detection} presents 
error detection results for \system{} and baselines across three benchmarks.
We find that \system{}
outperforms baselines in terms of
precision on both the Wikipedia
and the Excel benchmarks, despite
having a higher firing rate than
all but one baseline (T5). 
Auto-Detect, which is well-suited
to the type of mistakes present
in the Wikipedia dataset, performs
competitively. Overall, we find that error
detection is relatively
easier on the Wikipedia
benchmark, where tables
on average have fewer
rows, compared to the Excel
benchmark.  

On our synthetic benchmark, 
we find \system{} achieves the highest recall followed by
GPT-3.5. The learning-based
approach taken by Raha
also results in a recall
rate that is comparable
to the more expensive
GPT-3.5-based solution.

When performing qualitative
inspection of the errors detected,
we find that GPT-3.5 
can identify errors in semantic
substrings well.
For example, in the following column of financial quarters,
\emph{\{Q1-22,\; Q4-21,\; Q5-20,\; Q2-20,\; Q1-21\}}
GPT-3.5 correctly identifies the outlier to be \textit{Q5-20}.
However, GPT-3.5 fails
to detect syntactic errors
like \emph{S1.4} in the column,
\emph{\{S.1.2,\; S.2.3,\; S1.4,\; S.1.3,\; S.2.1\}},
where \emph{S1.4} is
missing a period
after \emph{S}.
Neural models like GPT-3.5 struggle at recognizing these patterns whereas \system{} can detect such errors using regular-expression-based patterns.

Other tools, like
Auto-Detect, work well on 
syntactic errors but fails
on semantic repairs. For example, consider the column of county and a numeric ID separated by a hyphen \emph{\{Alpine\_231,\; Kings\_721,\; Lake\_201,\; Santa Clara\_246,\; Nevad210\}} the correct repair here is \emph{Nevad210 $\rightarrow$ Nevada\_210}. This error involves a combination of syntactic and semantic inconsistency which most baseline systems struggle with. \system{} combines semantic information via masking into its pattern based syntactic repair engine and thus 
detects this error (and generates the correct repair).

\begin{table}[tb]
    \centering
    \caption{Table showing repair precision as the percentage of errors that \system{} and baseline systems can repair correctly out of the errors that were correctly detected by each system. \system{} has the highest repair rate compared to baseline systems.
    }
    \label{tab:repair_correct}
    \begin{tabularx}{\columnwidth}{Xccc}
        \toprule
        \textbf{System} & \textbf{Wikipedia} & \textbf{Excel} & \textbf{Synthetic} \\ \midrule
        WMRR & 87.3 & 89.9 & 78.2 \\
        HoloClean & 87.1 & 90.5 & 79.3 \\ \cmidrule(lr){1-4}
        Raha + GPT-3.5 & 85.0 & 85.0 & 76.0 \\
        Potter's-Wheel + GPT-3.5 & 84.9 & -- & -- \\
        Auto-Detect + GPT-3.5 & 85.2 & -- & -- \\ \cmidrule(lr){1-4}
        T5 & 67.4 & 70.1 & 68.8 \\
        GPT-3.5 & 86.5 & 86.3 & 76.3 \\ \cmidrule(lr){1-4}
        \system{} & \textbf{89.0} & \textbf{91.2} & \textbf{80.3} \\
        \bottomrule
    \end{tabularx}
\end{table}

\begin{figure}[tbh]
\centering
\includegraphics[width=0.95\columnwidth]{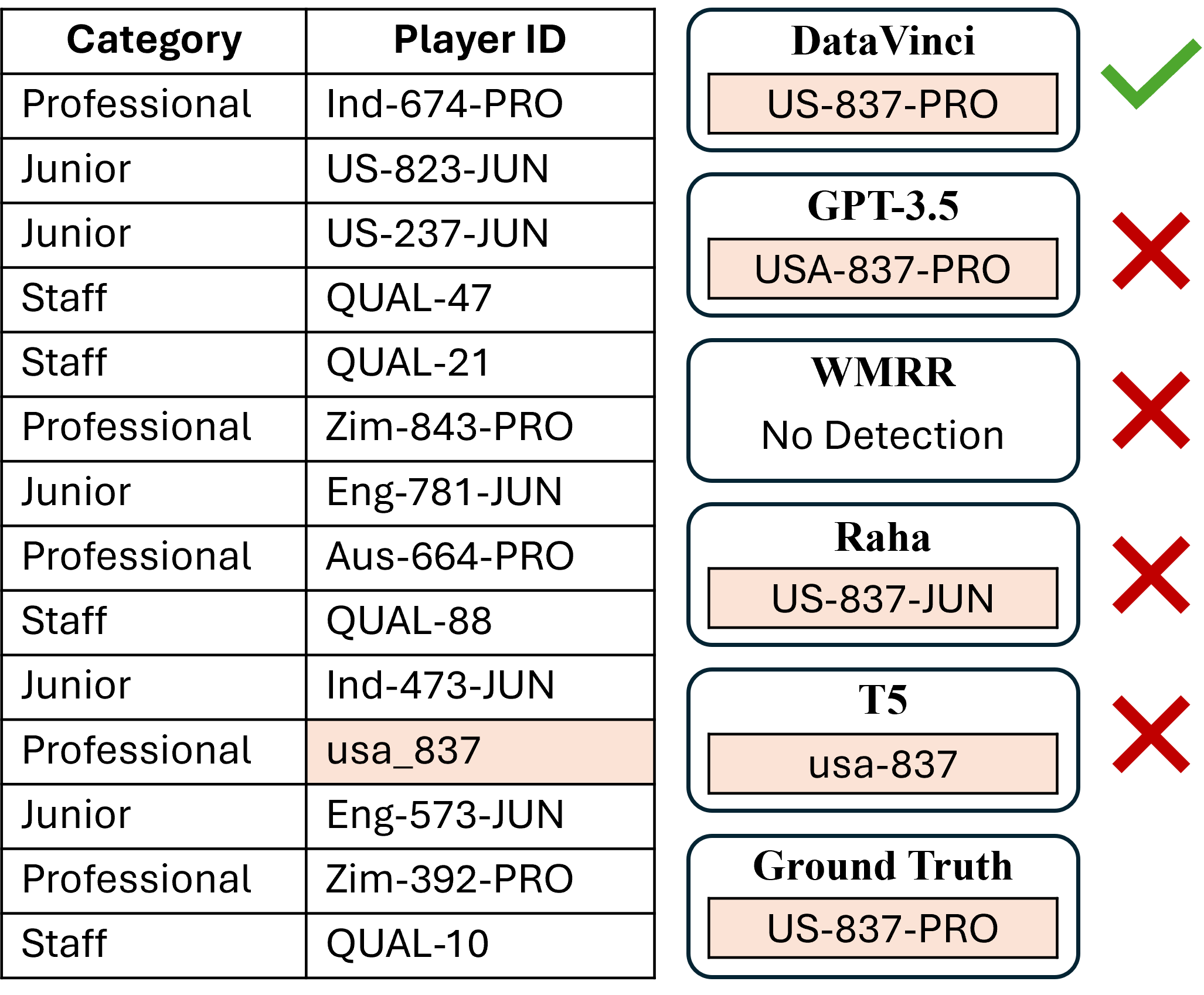}
\caption{Example from Excel benchmarks where \system{} generates the correct repair while baseline systems fail. The ground truth is \texttt{US-837-PRO}. \system{} combines semantic and syntactic substrings in its pattern and repairs the column correctly. Analysis for this example is presented in Figure~\ref{fig:arch}.
}
\label{fig:success_example}
\end{figure}

\subsection{Error Repair (RQ2)}

For repair, we find that often there are
cases where various possible data repairs
are reasonable and the correct repair cannot be uniquely identified. To account for this, we 
annotate repair suggestions as \emph{possible}
if this is the case, and then annotate the 
suggestion as correct (i.e. reasonable) or
not. In our results, we report
repair precision for certain cases and precision for \emph{possible}
cases separately for completeness.

Table~\ref{tab:repair}
shows the performance of \system{} and baselines systems for
repairing data on our benchmarks.
Note that repair metrics combine:
(1) detection (as a system must identify
an error to fix it), and (2)
whether the repair matches the
ground-truth repair.

We find that \system{} outperforms all baselines in terms of both certain and possible repairs on Wikipedia and Excel benchmarks and has the highest precision, recall and F1 score on the synthetic test set. Raha+GPT-3.5 and Auto-Detect+GPT-3.5 have high precision (Wikipedia), but we find that they have different
behaviors. Specifically,
Auto-Detect (by design)
does not support
inter-column dependencies,
while Raha struggles
to detect intra-column
patterns. WMRR and Potter's-Wheel capture both inter- and intra-column dependencies well but struggle with semantic repairs as they do not detect these issues.

Both GPT-3.5 and T5 perform significantly worse on the synthetic dataset as our noise operations predominantly introduced syntactic errors with minimal semantic content.

Table~\ref{tab:repair_correct}
shows the precision rates when
we only consider
correctly detected errors
for each system as a way to
disentangle the detection 
and correction effectiveness
of each system. We find that
repair precision is substantially
higher across the board, if 
we only consider correct detections,
and \system{} outperforms in
all three benchmark sets.

We look at cases where \system{} is able to accurately repair data errors while baselines fail. We find that these are mostly where either (1) \system{} is able to leverage its semantic masking to suppress false positives; or (2) \system{} detects a semantic anomaly using patterns. Figure~\ref{fig:success_example} shows an example from the Excel benchmarks, which contains a tournament table having columns category (Junior or Professional) and Player-ID which has three components (Country code, unique numeric ID, first three letters of category). For non competing players, the ID is QUAL- followed by a unique numeric ID. \emph{usa\_837} is an outlier in the Player-ID column and the correct repair should change it to \emph{US-837-PRO}. As highlighted above, \system{} utilizes (1) semantic masking to repair \emph{usa $\rightarrow$ US} and uses (2) patterns, paired with concretization value constraints, to detect that the category substring is \emph{PRO}.

\begin{figure}[tb]
\centering
\includegraphics[width=\columnwidth]{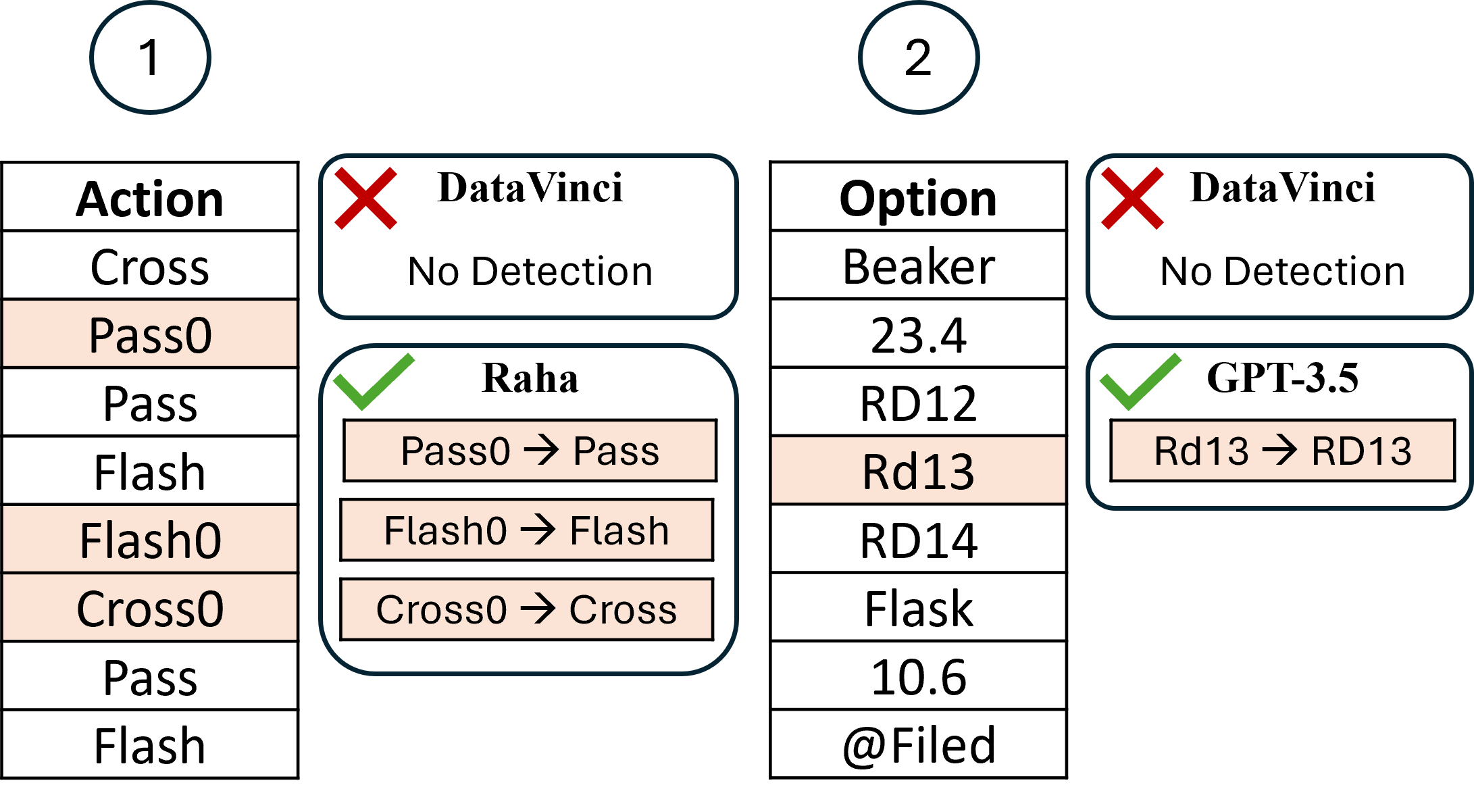}
\caption{Example from Excel benchmarks where \system{} fails to generate the correct repair but a baseline system succeeds. Error cells are highlighted. We show all repairs (denoted by $\rightarrow$) suggested. \system{} is unable to detect these errors because they either \textcircled{1} satisfy a \significant{} pattern, or \textcircled{2} the column has irregular data and no \significant{} pattern is learned.
}
\label{fig:failure_example}
\end{figure}

We also look at cases where \system{} failed to generate the correct repair but one of the baseline systems succeeded. 
We find that these cases were mostly the result of either 
(1) the column does not have any \significant{} patterns due to irregular data, or 
(2) the error rate is too high and as a result the outlier is covered by a \significant{} pattern.
Figure~\ref{fig:failure_example} shows one example for each case from the Excel benchmarks along with the incorrect repair generated by \system{} and the correct repair generated by the baseline.
In the first example, \system{} learns two \significant{} patterns $R_M = $ \{\texttt{[A-Z]+}, \texttt{[A-Z]+0}\} and hence, does not detect the error. In the second example, since the column contains irregular data, \system{} is unable to learn a \significant{} pattern $R_M = \{\}$ and does not detect any errors.

\subsection{Execution-Guidance (RQ3)}

\begin{table}[]
\centering
\caption{
Execution success rates at formula and cell-level after applying repair suggestions for each system on our Excel Formulas
benchmark. 
We split out formulas that depend on single and multiple columns. 
We report the \emph{No Repair} (i.e. starting point) success rates first for comparison.
We do not report HoloClean as it is expensive to run and did not complete in the 24 hours time limit.
Applying \system{}'s execution-guided repairs leads to more successful executions. 
}
\begin{tabularx}{\columnwidth}{Xcccc}
\toprule
 & \multicolumn{2}{c}{\textbf{\small Single Column}} & \multicolumn{2}{c}{\textbf{\small Multi Column}}  \\ \midrule
\textbf{\small Type} & \textbf{\small Formula} & \textbf{\small Cell} & \textbf{\small Formula} & \textbf{\small Cell}  \\ \midrule
\small No Repair & 0.0\% & 85.8\% & 0.0\% & 81.4\%  \\
\small WMRR & 32.6\% & 94.4\% & 29.6\% & 90.1\%  \\
\small Raha + GPT-3.5 & 34.5\% & 92.6\% & 31.4\% & 88.3\%  \\
\small T5 & 11.2\% & 89.4\% & 6.4\% & 86.2\%  \\
\small\system{} Unsupervised & 43.2\% & 94.3\% & 35.7\% & 90.9\%  \\
\small\system{}+Execution & 54.0\% & 96.5\% & 47.8\% & 94.0\%  \\ \bottomrule
\end{tabularx}
\label{tab:execution-guided}
\end{table}

We use our Excel Formula
benchmark to evaluate the extent
to which \system{}
can use execution information
to provide improved repairs.
We report two execution metrics: the fraction 
of cells that no longer result in 
an error value, as well as the fraction of columns where no cell results in an error value (i.e., the formula succeeds fully).

To carry out this experiment, 
we run all baselines, apply
repair suggestions \emph{only}
on values that are an input
into a row that has an
error value when the formula
is originally executed.
We report formula-level and cell-level
successful execution rates
after applying each systems'
suggestions.

Table~\ref{tab:execution-guided} summarizes our results. We find that
\system{} with execution-guided learning
improves over all baselines
\footnote{we exclude HoloClean as it did not scale to this task. We let HoloClean run for 24 hours and it only covered 24\% of the formula benchmarks.}
(including the
fully unsupervised \system{}).
While all systems have implicit access
to execution information, since we only apply their repairs to inputs associated with erroneous executions, \system{} with execution-guided learning is the only system
that also incorporates this information (by affecting the patterns learned: $R_m$) when learning how to
repair the data error.

\begin{figure}[tb]
    \captionsetup{justification=centering}
        \centering
        \begin{subfigure}{.50\columnwidth}
            \centering
            \includegraphics[width=\columnwidth]{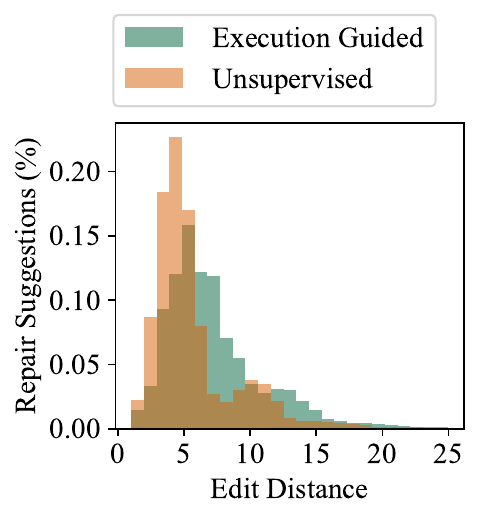}
            \caption{Edit Distance}
            \label{fig:accuracyLength}
        \end{subfigure}
        \begin{subfigure}{.48\columnwidth}
            \includegraphics[width=\columnwidth]{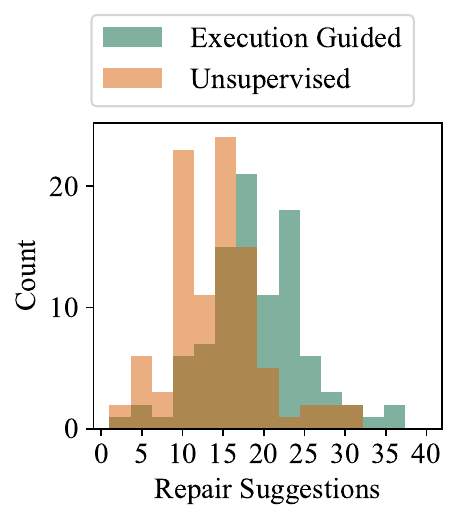}
            \caption{Number of Repairs}
            \label{fig:silhouetteLength}
        \end{subfigure}
    \captionsetup{justification=justified}
        \caption{
        Comparison of unsupervised and execution-guided
        \system{}. Both the (a)
        distribution of original value to suggested repair edit distances and (b) number of repairs per column
        move higher when given execution information.
        Jointly, this suggests that with execution information
        \system{} can offer more repairs, with higher complexity (as proxied by string edit distance).
        }
        \label{fig:exec-unsupervised-diffs}
\end{figure}

\begin{figure}[tbh]
\centering
\includegraphics[width=\columnwidth]{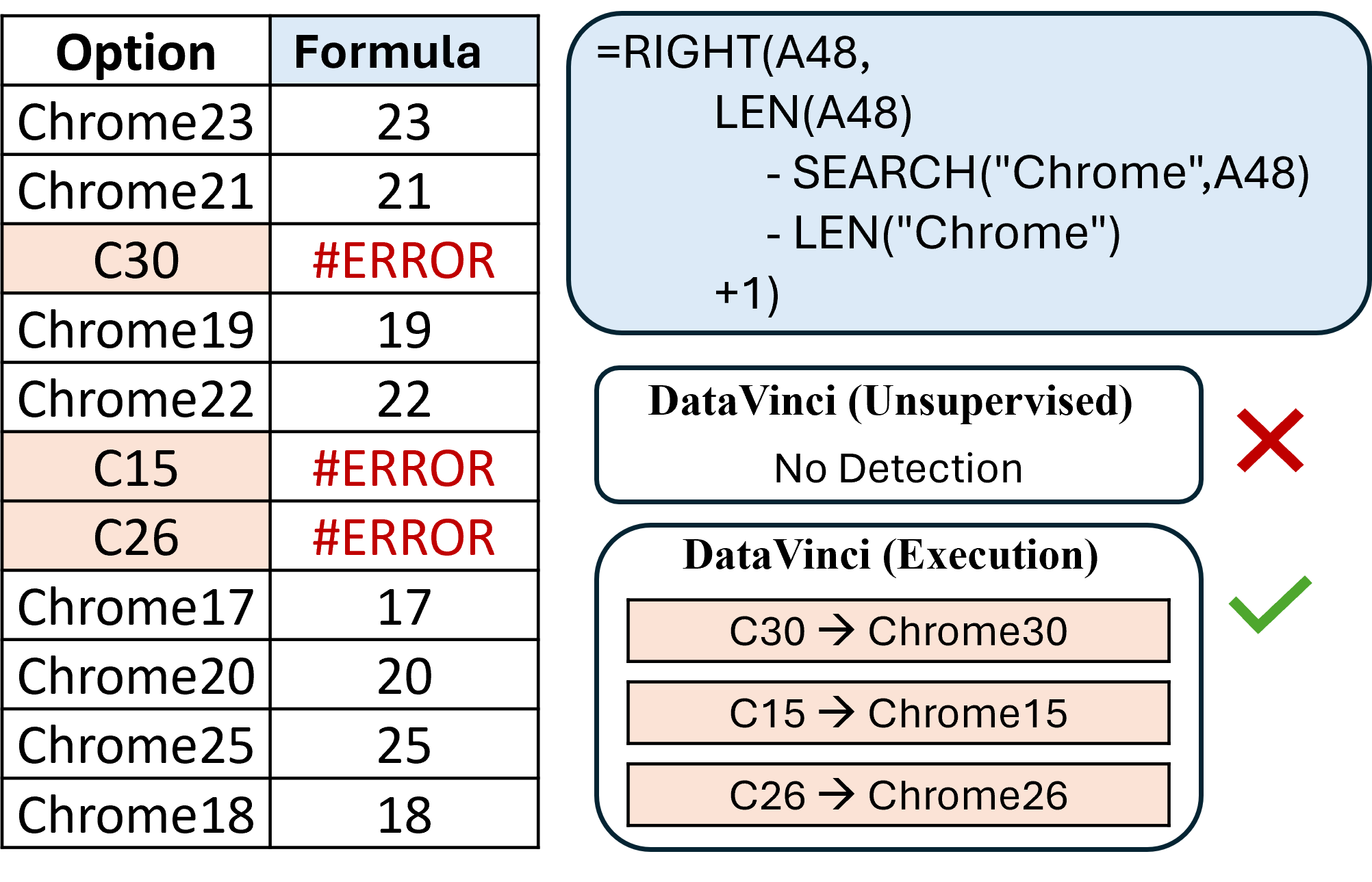}
\caption{Example from Excel Formulas benchmark where \system{} with execution-guided repair can provide repair suggestions that lead to successful formula execution but the unsupervised variant cannot because the outlier pattern, \texttt{C[0-9]\{2\}} occurs frequent enough to be considered a significant pattern. Error output values
are shown in red font, and
data errors in input values
are highlighted.
}
\label{fig:execution_example}
\end{figure}

Figure~\ref{fig:exec-unsupervised-diffs} shows that
the distribution of string edit distances from original data error to repaired suggestion, as well as the number of repairs
per column. We find that when provided with execution
information \system{} can produce more repairs, and these
repairs tend to have a higher distance to the original value
(implying possibly more complex repairs). For example,
Figure~\ref{fig:execution_example} shows an example where \system{} with execution learns the correct repair resulting in successful execution of the formula while the unsupervised variant is unable to provide any suggestions because the pattern \texttt{C[0-9]\{2\}} repeats enough times to be considered as a significant pattern.

\subsection{Design Decision (RQ4)}

We study how the various design decisions impact the performance of \system{}. Namely,
we investigate
the importance of
semantic abstraction/concretization,
concretization value constraints,
and ranking. We carry out experiments on 
our synthetically corrupted benchmark.
We summarize
our results in 
Table~\ref{tab:ablations}.

\begin{table}[tb]
\centering
\caption{
Repair precision, recall and F1 score 
on our synthetically corrupted
benchmark for
different \system{}
ablations. Full
\system{} outperforms all ablations.
Removing semantic abstraction
and removing learned concretization constraints (which concretize
abstract edits)
have the most impact on F1 score.
}
\label{tab:ablations}
\begin{tabularx}{\columnwidth}{Xccc}
\toprule
\textbf{Model} & \textbf{Precision} & \textbf{Recall} & \textbf{F1} \\ \midrule
No semantic abstraction & 50.3 & 62.9 & 55.9 \\
Limited semantic concretization & 52.0 & 65.6 & 58.0 \\ \cmidrule(lr){1-4}
No learned concretization & 46.3 & 51.0 & 48.5 \\ 
Edit distance ranking & 53.2 & 67.1 & 69.3 \\ \cmidrule(lr){1-4}
\system{} & \textbf{54.1} & \textbf{68.9} & \textbf{60.6} \\
\bottomrule
\end{tabularx}
\end{table}

\subsubsection{Semantic Substrings}

To evaluate the impact of having semantic information in repairs (see Section~\ref{sec:semantic}), we evaluate two versions of \system{}.
We implement a version of \system{}
that 
treats all strings as purely syntactic
(\emph{No semantic abstraction})
and a version that can perform
semantic abstraction but is
restricted to re-use the same
substring for concretization, 
meaning it can not repair errors in 
semantic substrings (\emph{Limited semantic concretization}).
We find that both versions
result in a lower performance
compared to full \system{}, but removing
semantics altogether has a comparatively larger
impact on precision and F1.

\subsubsection{Concretization and Ranking}

To study the effect of learned concretization value constraints and ranking (see Section~\ref{ssec:concretization} and ~\ref{sec:ranking}), we design two ablated version of \system{}, (1) where \system{} does not learn to concretize abstract edits, instead enumerates all candidates and directly passes them through to the heuristic ranker (\emph{No learned concretization}); and (2) a version
where all candidates are
ranked just based on the shortest string edit distance with respect to the original data error (\emph{Edit distance ranking}). Table~\ref{tab:ablations} shows that while removing
either learned concretization or ranking has a negative impact on performance, 
removing learned concretization constraints has a larger
effect.

\subsection{Runtime Performance}

\begin{table}[tb]
    \centering
    \caption{
    Comparing time (milliseconds), 
    disk space (MB), and GPU plus CPU memory used (MB), averaged
    over Wikipedia benchmarks. (*) denotes systems which we did not run and the reported stats are the ones published by Auto-Detect authors for this dataset. Raha/Potter's Wheel/Auto-Detect only includes the detection time as repair is performed by GPT head in our
    experiments. GPT-3.5 was used via API and the reported time includes network latency.
 }
    \label{tab:time-memory}
    \begin{tabularx}{\columnwidth}{Xccc}
        \toprule
        \textbf{System} & \textbf{Time(ms)} & \textbf{Disk(MB)} & \textbf{Memory(MB)} \\ \midrule
        WMRR & 247.4 & 4.6 & 914.5 \\
        HoloClean & 1049.3 & 996.3 & 1647.2 \\ \cmidrule(lr){1-4}
        Raha & 321.8 & 65.3 & 645.4 \\ 
        Potter's-Wheel* & 110.0 & - & - \\
        Auto-Detect* & 290.0 & - & - \\ \cmidrule(lr){1-4}
        T5 & 858.3 & 886.2 & 1534.2 \\
        GPT-3.5 & 1325.6 & - & - \\ \cmidrule(lr){1-4}
        {\system{}} & {261.5} & {5.6} & {10.5} \\
        \bottomrule
    \end{tabularx}
\end{table}

Table~\ref{tab:time-memory} shows the average time taken, disk space used and 
RAM + GPU VRAM used to detect (Raha, Potter's Wheel, Auto-Detect) or detect+repair (WMRR, HoloClean, T5, GPT-3.5, \system{}) errors on the Wikipedia benchmark.
We report Potter's Wheel and Auto-Detect
based on the Auto-Detect paper \cite{autodetect} which reports these metrics in a similar environment and on the same benchmarks.
We note that GPT-3.5 includes network
time. We find that in terms
of time and disk space, \system{} is competitive
with alternatives such as WMRR, Raha, and Auto-Detect, while using substantially less
RAM.
Since, T5 and HoloClean also utilize a GPU and don't run purely on CPU memory we report the sum of GPU and CPU memory usage under Memory.
We only report inference resources but it is worth noting that T5 also requires training which took 4 hours on a K80 GPU.

HoloClean and T5 are the most resource intensive systems, as a result of their implementation complexity. \system{}, WMRR and Raha are up to 4 times faster than HoloClean/GPT-3.5 and require fewer resources to run.

\section{Limitations}
We describe some limitations of \system{}. 
We have only evaluated \system{}
on English language values and applicability
to non-English datasets may be limited.
\system{} does not handle inter-table
constraints, limiting its effectiveness 
when data consistency relies on relationships across multiple tables. \system{} relies on identifying recurring patterns to 
perform error detection/repair, which may
not occur in all data. Execution-guided
repair may mitigate this limitation but
its applicability is limited by the availability of programs
that read the target data
and the ability to easily execute these programs.
\section{Related Work}

Data error detection
has been the subject
of active investigation
in the
data management community.
Prior work has developed systems that can
identify qualitative data
errors (e.g. incorrect
value structure) or
quantitative data errors
(e.g. unlikely values
given a distribution). Often systems further make distinctions between syntactic issues and semantic data errors~\cite{raha}.
The seminal Potter's 
Wheel~\cite{raman2001potter}
system introduced
an interactive data cleaning
system where users employ
a spreadsheet-like environment
to annotate data errors with corrections, which the system
then learns to apply throughout
the data. More recent
work like AutoDetect~\cite{autodetect}
uses large-scale
co-occurrence statistics,
along with pattern-based
generalization, to 
achieve high-precision
error detection. Raha~\cite{origraha} achieves configuration-free error detection by combining
many error strategies, clustering data based on these strategies' annotations, efficiently
gathering user feedback on possible
errors from these clusters, and then training
a model on this feedback to detect errors throughout the dataset.
Commercial platforms like
Trifacta~\cite{trifacta}
and PowerBI~\cite{microsoft_clean_data_power_bi} typically offer data cleaning based on a fixed set of patterns or
common data issues (e.g. leading spaces).
In contrast
to this line of work,
\system{} does not use fixed patterns to detect errors but rather learns them. \system{} does not only
detect data errors but
also repairs them.
In addition, \system{} can
detect and repair errors
in strings with
both syntactic and semantic
substrings.

While detecting data
errors can help users identify issues in their dataset,
correcting these errors can also be costly. As a result, past work has explored
not only detecting but also repairing errors identified. 
HoloClean~\cite{holoclean}  
allows users to (optionally) specify denial constraints, which it combines with 
error detectors, to build a probablistic data cleaning
program, which unifies these
different signals.
WMRR~\cite{wmrr} presents
an unsupervised approach to
learning cleaning rules,
removing the need for users
to specify denial constraints
or resolution rules.
ActiveClean~\cite{krishnan2016activeclean} and Baran~\cite{mahdavi2020baran,raha} are semi-supervised tools to repair data based on few user examples. With the exception of WMRR (which our evaluation shows achieves lower detection/repair performance), this line of work requires a human in the loop.
Like this work, \system{}
can provide repair suggestions
for errors detected. 
In contrast to these systems,
\system{} uses a pattern-based
approach to data repair,
does not require any user specification in the form of constraints or annotated examples,
and can repair strings
that have a combination of
both syntactic and semantic
substrings.

Transformer models \cite{transformer} have recently gained a lot of popularity and language \cite{T5} and code tasks \cite{codet5}. As part of our evaluation, we employ T5 and
GPT-3.5 as baselines to perform data error detection
and repair. Prior work~\cite{narayan2022can} has explored using
foundation models for data management tasks including data validation and cleaning.

\system{} can repair strings that contain both syntactic
and semantic substrings, as described in Section~\ref{sec:semantic}.
FlashGPT~\cite{verbruggen2021semantic} showed that
LLM-based semantic transformations can be integrated into
a programming-by-example synthesizer that learns
syntactic string transformations.
More recently, SMORE~\cite{chen2023data} presented
a formalization of semantic regular expressions,
which generalize traditional regular expressions to 
include semantic substrings. SMORE can learn such
semantic regexes given positive/negative examples.
Potter's Wheel~\cite{raman2001potter}
anticipated the combination of
semantic substrings and syntactic substrings for data cleaning, and allowed users to define 
membership tests associated
with semantic types, and used
these definitions to guide its transformations.
Like this line of work, \system{} combines
syntactic and semantic information. Like SMORE,
\system{} leverages regular expressions to formally
describe string values. In contrast to SMORE, FlashGPT, and Potter's Wheel,
\system{} does not have 
user examples or definitions
but
rather learns patterns fully unsupervised. Like SMORE,
\system{} uses an LLM to identify semantic substrings, but
when learning the regular expression \system{} employs
abstraction/concretization, which allows it to use
an existing regular expression learner~\cite{padhi2018flashprofile}.

\section{Conclusion}
In this paper we propose \system{}, a tool for automatic repair of string data errors. 
\system{} learns majority patterns over columns and uses these to detect string data errors. 
Because errors may occur in strings with both
syntactic and semantic substrings, \system{}
employs an LLM to mask semantic substrings before
learning a pattern.
\system{} suggests repairs for the detected errors by deriving minimal edits to the data error that
lead to satisfying a majority pattern.
Because majority patterns may not always occur
or capture errors, \system{} can address
this challenge by incorporating program
execution information.
We evaluate \system{} against 7 baselines on four
existing and new benchmarks and show \system{}
achieves higher detection and repair performance.
We release 
scripts to reproduce
our novel benchmarks for
future data cleaning research.

\section{Acknowledgments}
We would like to thank Yair Helman, Einam Schonberg, Tal Kariv, Noa Feiger, Israela Solomon, Irena Berezovsky, Guy Hunkin, David Schwartz, Danielle Maor, Danielle Fainman, Tsofiya Aiello, Jack Williams and Andy Gordon
for their feedback on this research.

\bibliographystyle{ACM-Reference-Format}
\bibliography{references}


\begin{thebibliography}{27}


\ifx \showCODEN    \undefined \def \showCODEN     #1{\unskip}     \fi
\ifx \showDOI      \undefined \def \showDOI       #1{#1}\fi
\ifx \showISBNx    \undefined \def \showISBNx     #1{\unskip}     \fi
\ifx \showISBNxiii \undefined \def \showISBNxiii  #1{\unskip}     \fi
\ifx \showISSN     \undefined \def \showISSN      #1{\unskip}     \fi
\ifx \showLCCN     \undefined \def \showLCCN      #1{\unskip}     \fi
\ifx \shownote     \undefined \def \shownote      #1{#1}          \fi
\ifx \showarticletitle \undefined \def \showarticletitle #1{#1}   \fi
\ifx \showURL      \undefined \def \showURL       {\relax}        \fi
\providecommand\bibfield[2]{#2}
\providecommand\bibinfo[2]{#2}
\providecommand\natexlab[1]{#1}
\providecommand\showeprint[2][]{arXiv:#2}

\bibitem[\protect\citeauthoryear{??}{tri}{[n.d.]}]%
        {trifacta}
 \bibinfo{year}{[n.d.]}\natexlab{}.
\newblock \bibinfo{title}{{Trifacta: Data Cleansing Tool}}.
\newblock
  \bibinfo{howpublished}{\url{https://docs.trifacta.com/display/AAC/Data+Cleansing+Tool}}.
\newblock
\newblock
\shownote{Accessed: \today.}


\bibitem[\protect\citeauthoryear{Abu~Ahmad and Wang}{Abu~Ahmad and
  Wang}{2020}]%
        {wmrr}
\bibfield{author}{\bibinfo{person}{Hiba Abu~Ahmad} {and}
  \bibinfo{person}{Hongzhi Wang}.} \bibinfo{year}{2020}\natexlab{}.
\newblock \showarticletitle{Automatic weighted matching rectifying rule
  discovery for data repairing: Can we discover effective repairing rules
  automatically from dirty data?}
\newblock \bibinfo{journal}{\emph{The VLDB Journal}} \bibinfo{volume}{29},
  \bibinfo{number}{6} (\bibinfo{year}{2020}), \bibinfo{pages}{1433--1447}.
\newblock


\bibitem[\protect\citeauthoryear{Brown, Mann, Ryder, Subbiah, Kaplan, Dhariwal,
  Neelakantan, Shyam, Sastry, Askell, Agarwal, Herbert-Voss, Krueger, Henighan,
  Child, Ramesh, Ziegler, Wu, Winter, Hesse, Chen, Sigler, Litwin, Gray, Chess,
  Clark, Berner, McCandlish, Radford, Sutskever, and Amodei}{Brown
  et~al\mbox{.}}{2020}]%
        {gpt3}
\bibfield{author}{\bibinfo{person}{Tom Brown}, \bibinfo{person}{Benjamin Mann},
  \bibinfo{person}{Nick Ryder}, \bibinfo{person}{Melanie Subbiah},
  \bibinfo{person}{Jared~D Kaplan}, \bibinfo{person}{Prafulla Dhariwal},
  \bibinfo{person}{Arvind Neelakantan}, \bibinfo{person}{Pranav Shyam},
  \bibinfo{person}{Girish Sastry}, \bibinfo{person}{Amanda Askell},
  \bibinfo{person}{Sandhini Agarwal}, \bibinfo{person}{Ariel Herbert-Voss},
  \bibinfo{person}{Gretchen Krueger}, \bibinfo{person}{Tom Henighan},
  \bibinfo{person}{Rewon Child}, \bibinfo{person}{Aditya Ramesh},
  \bibinfo{person}{Daniel Ziegler}, \bibinfo{person}{Jeffrey Wu},
  \bibinfo{person}{Clemens Winter}, \bibinfo{person}{Chris Hesse},
  \bibinfo{person}{Mark Chen}, \bibinfo{person}{Eric Sigler},
  \bibinfo{person}{Mateusz Litwin}, \bibinfo{person}{Scott Gray},
  \bibinfo{person}{Benjamin Chess}, \bibinfo{person}{Jack Clark},
  \bibinfo{person}{Christopher Berner}, \bibinfo{person}{Sam McCandlish},
  \bibinfo{person}{Alec Radford}, \bibinfo{person}{Ilya Sutskever}, {and}
  \bibinfo{person}{Dario Amodei}.} \bibinfo{year}{2020}\natexlab{}.
\newblock \showarticletitle{Language Models are Few-Shot Learners}. In
  \bibinfo{booktitle}{\emph{Advances in Neural Information Processing
  Systems}}, \bibfield{editor}{\bibinfo{person}{H.~Larochelle},
  \bibinfo{person}{M.~Ranzato}, \bibinfo{person}{R.~Hadsell},
  \bibinfo{person}{M.F. Balcan}, {and} \bibinfo{person}{H.~Lin}} (Eds.),
  Vol.~\bibinfo{volume}{33}. \bibinfo{publisher}{Curran Associates, Inc.},
  \bibinfo{pages}{1877--1901}.
\newblock
\urldef\tempurl%
\url{https://proceedings.neurips.cc/paper/2020/file/1457c0d6bfcb4967418bfb8ac142f64a-Paper.pdf}
\showURL{%
\tempurl}


\bibitem[\protect\citeauthoryear{Chen, Banerjee, Demiralp, Durrett, and
  Dillig}{Chen et~al\mbox{.}}{2023}]%
        {chen2023data}
\bibfield{author}{\bibinfo{person}{Qiaochu Chen}, \bibinfo{person}{Arko
  Banerjee}, \bibinfo{person}{{\c{C}}a{\u{g}}atay Demiralp},
  \bibinfo{person}{Greg Durrett}, {and} \bibinfo{person}{Isil Dillig}.}
  \bibinfo{year}{2023}\natexlab{}.
\newblock \showarticletitle{Data Extraction via Semantic Regular Expression
  Synthesis}.
\newblock \bibinfo{journal}{\emph{arXiv preprint arXiv:2305.10401}}
  (\bibinfo{year}{2023}).
\newblock


\bibitem[\protect\citeauthoryear{Chu, Ilyas, Krishnan, and Wang}{Chu
  et~al\mbox{.}}{2016}]%
        {chu2016data}
\bibfield{author}{\bibinfo{person}{Xu Chu}, \bibinfo{person}{Ihab~F Ilyas},
  \bibinfo{person}{Sanjay Krishnan}, {and} \bibinfo{person}{Jiannan Wang}.}
  \bibinfo{year}{2016}\natexlab{}.
\newblock \showarticletitle{Data cleaning: Overview and emerging challenges}.
  In \bibinfo{booktitle}{\emph{Proceedings of the 2016 international conference
  on management of data}}. \bibinfo{pages}{2201--2206}.
\newblock


\bibitem[\protect\citeauthoryear{Heidari, McGrath, Ilyas, and
  Rekatsinas}{Heidari et~al\mbox{.}}{2019}]%
        {holodetect}
\bibfield{author}{\bibinfo{person}{Alireza Heidari}, \bibinfo{person}{Joshua
  McGrath}, \bibinfo{person}{Ihab~F Ilyas}, {and} \bibinfo{person}{Theodoros
  Rekatsinas}.} \bibinfo{year}{2019}\natexlab{}.
\newblock \showarticletitle{Holodetect: Few-shot learning for error detection}.
  In \bibinfo{booktitle}{\emph{Proceedings of the 2019 International Conference
  on Management of Data}}. \bibinfo{pages}{829--846}.
\newblock


\bibitem[\protect\citeauthoryear{Huang and He}{Huang and He}{2018}]%
        {autodetect}
\bibfield{author}{\bibinfo{person}{Zhipeng Huang} {and} \bibinfo{person}{Yeye
  He}.} \bibinfo{year}{2018}\natexlab{}.
\newblock \showarticletitle{Auto-detect: Data-driven error detection in
  tables}. In \bibinfo{booktitle}{\emph{Proceedings of the 2018 International
  Conference on Management of Data}}. \bibinfo{pages}{1377--1392}.
\newblock


\bibitem[\protect\citeauthoryear{Hulsebos, Hu, Bakker, Zgraggen, Satyanarayan,
  Kraska, Demiralp, and Hidalgo}{Hulsebos et~al\mbox{.}}{2019}]%
        {sherlock}
\bibfield{author}{\bibinfo{person}{Madelon Hulsebos}, \bibinfo{person}{Kevin
  Hu}, \bibinfo{person}{Michiel Bakker}, \bibinfo{person}{Emanuel Zgraggen},
  \bibinfo{person}{Arvind Satyanarayan}, \bibinfo{person}{Tim Kraska},
  \bibinfo{person}{\c{C}agatay Demiralp}, {and} \bibinfo{person}{C{\'e}sar
  Hidalgo}.} \bibinfo{year}{2019}\natexlab{}.
\newblock \showarticletitle{Sherlock: A Deep Learning Approach to Semantic Data
  Type Detection}. In \bibinfo{booktitle}{\emph{Proceedings of the 25th ACM
  SIGKDD International Conference on Knowledge Discovery \&\#38; Data Mining}}.
  \bibinfo{publisher}{ACM}.
\newblock


\bibitem[\protect\citeauthoryear{Krishnan, Wang, Wu, Franklin, and
  Goldberg}{Krishnan et~al\mbox{.}}{2016}]%
        {krishnan2016activeclean}
\bibfield{author}{\bibinfo{person}{Sanjay Krishnan}, \bibinfo{person}{Jiannan
  Wang}, \bibinfo{person}{Eugene Wu}, \bibinfo{person}{Michael~J Franklin},
  {and} \bibinfo{person}{Ken Goldberg}.} \bibinfo{year}{2016}\natexlab{}.
\newblock \showarticletitle{Activeclean: Interactive data cleaning for
  statistical modeling}.
\newblock \bibinfo{journal}{\emph{Proceedings of the VLDB Endowment}}
  \bibinfo{volume}{9}, \bibinfo{number}{12} (\bibinfo{year}{2016}),
  \bibinfo{pages}{948--959}.
\newblock


\bibitem[\protect\citeauthoryear{Mahdavi and Abedjan}{Mahdavi and
  Abedjan}{2020}]%
        {mahdavi2020baran}
\bibfield{author}{\bibinfo{person}{Mohammad Mahdavi} {and}
  \bibinfo{person}{Ziawasch Abedjan}.} \bibinfo{year}{2020}\natexlab{}.
\newblock \showarticletitle{Baran: Effective error correction via a unified
  context representation and transfer learning}.
\newblock \bibinfo{journal}{\emph{Proceedings of the VLDB Endowment}}
  \bibinfo{volume}{13}, \bibinfo{number}{12} (\bibinfo{year}{2020}),
  \bibinfo{pages}{1948--1961}.
\newblock


\bibitem[\protect\citeauthoryear{Mahdavi and Abedjan}{Mahdavi and
  Abedjan}{2021}]%
        {raha}
\bibfield{author}{\bibinfo{person}{Mohammad Mahdavi} {and}
  \bibinfo{person}{Ziawasch Abedjan}.} \bibinfo{year}{2021}\natexlab{}.
\newblock \showarticletitle{Semi-Supervised Data Cleaning with Raha and
  Baran.}. In \bibinfo{booktitle}{\emph{CIDR}}.
\newblock


\bibitem[\protect\citeauthoryear{Mahdavi, Abedjan, Castro~Fernandez, Madden,
  Ouzzani, Stonebraker, and Tang}{Mahdavi et~al\mbox{.}}{2019}]%
        {origraha}
\bibfield{author}{\bibinfo{person}{Mohammad Mahdavi}, \bibinfo{person}{Ziawasch
  Abedjan}, \bibinfo{person}{Raul Castro~Fernandez}, \bibinfo{person}{Samuel
  Madden}, \bibinfo{person}{Mourad Ouzzani}, \bibinfo{person}{Michael
  Stonebraker}, {and} \bibinfo{person}{Nan Tang}.}
  \bibinfo{year}{2019}\natexlab{}.
\newblock \showarticletitle{Raha: A configuration-free error detection system}.
  In \bibinfo{booktitle}{\emph{Proceedings of the 2019 International Conference
  on Management of Data}}. \bibinfo{pages}{865--882}.
\newblock


\bibitem[\protect\citeauthoryear{Microsoft}{Microsoft}{[n.d.]}]%
        {microsoft_clean_data_power_bi}
\bibfield{author}{\bibinfo{person}{Microsoft}.}
  \bibinfo{year}{[n.d.]}\natexlab{}.
\newblock \bibinfo{booktitle}{\emph{{Clean, transform, and load data in Power
  BI}}}.
\newblock
\urldef\tempurl%
\url{https://learn.microsoft.com/en-us/training/modules/clean-data-power-bi/}
\showURL{%
\tempurl}
\newblock
\shownote{Accessed: \today.}


\bibitem[\protect\citeauthoryear{Narayan, Chami, Orr, Arora, and
  R{\'e}}{Narayan et~al\mbox{.}}{2022}]%
        {narayan2022can}
\bibfield{author}{\bibinfo{person}{Avanika Narayan}, \bibinfo{person}{Ines
  Chami}, \bibinfo{person}{Laurel Orr}, \bibinfo{person}{Simran Arora}, {and}
  \bibinfo{person}{Christopher R{\'e}}.} \bibinfo{year}{2022}\natexlab{}.
\newblock \showarticletitle{Can Foundation Models Wrangle Your Data?}
\newblock \bibinfo{journal}{\emph{arXiv preprint arXiv:2205.09911}}
  (\bibinfo{year}{2022}).
\newblock


\bibitem[\protect\citeauthoryear{Padhi, Jain, Perelman, Polozov, Gulwani, and
  Millstein}{Padhi et~al\mbox{.}}{2018}]%
        {padhi2018flashprofile}
\bibfield{author}{\bibinfo{person}{Saswat Padhi}, \bibinfo{person}{Prateek
  Jain}, \bibinfo{person}{Daniel Perelman}, \bibinfo{person}{Oleksandr
  Polozov}, \bibinfo{person}{Sumit Gulwani}, {and} \bibinfo{person}{Todd
  Millstein}.} \bibinfo{year}{2018}\natexlab{}.
\newblock \showarticletitle{FlashProfile: a framework for synthesizing data
  profiles}.
\newblock \bibinfo{journal}{\emph{Proceedings of the ACM on Programming
  Languages}} \bibinfo{volume}{2}, \bibinfo{number}{OOPSLA}
  (\bibinfo{year}{2018}), \bibinfo{pages}{1--28}.
\newblock


\bibitem[\protect\citeauthoryear{Qahtan, Tang, Ouzzani, Cao, and
  Stonebraker}{Qahtan et~al\mbox{.}}{2020}]%
        {pfd}
\bibfield{author}{\bibinfo{person}{Abdulhakim Qahtan}, \bibinfo{person}{Nan
  Tang}, \bibinfo{person}{Mourad Ouzzani}, \bibinfo{person}{Yang Cao}, {and}
  \bibinfo{person}{Michael Stonebraker}.} \bibinfo{year}{2020}\natexlab{}.
\newblock \showarticletitle{Pattern functional dependencies for data cleaning}.
\newblock  (\bibinfo{year}{2020}).
\newblock


\bibitem[\protect\citeauthoryear{Raffel, Shazeer, Roberts, Lee, Narang, Matena,
  Zhou, Li, and Liu}{Raffel et~al\mbox{.}}{2020}]%
        {T5}
\bibfield{author}{\bibinfo{person}{Colin Raffel}, \bibinfo{person}{Noam
  Shazeer}, \bibinfo{person}{Adam Roberts}, \bibinfo{person}{Katherine Lee},
  \bibinfo{person}{Sharan Narang}, \bibinfo{person}{Michael Matena},
  \bibinfo{person}{Yanqi Zhou}, \bibinfo{person}{Wei Li}, {and}
  \bibinfo{person}{Peter~J. Liu}.} \bibinfo{year}{2020}\natexlab{}.
\newblock \showarticletitle{Exploring the Limits of Transfer Learning with a
  Unified Text-to-Text Transformer}.
\newblock \bibinfo{journal}{\emph{Journal of Machine Learning Research}}
  \bibinfo{volume}{21}, \bibinfo{number}{140} (\bibinfo{year}{2020}),
  \bibinfo{pages}{1--67}.
\newblock
\urldef\tempurl%
\url{http://jmlr.org/papers/v21/20-074.html}
\showURL{%
\tempurl}


\bibitem[\protect\citeauthoryear{Raman and Hellerstein}{Raman and
  Hellerstein}{2001}]%
        {raman2001potter}
\bibfield{author}{\bibinfo{person}{Vijayshankar Raman} {and}
  \bibinfo{person}{Joseph~M Hellerstein}.} \bibinfo{year}{2001}\natexlab{}.
\newblock \showarticletitle{Potter's wheel: An interactive data cleaning
  system}. In \bibinfo{booktitle}{\emph{VLDB}}, Vol.~\bibinfo{volume}{1}.
  \bibinfo{pages}{381--390}.
\newblock


\bibitem[\protect\citeauthoryear{Rekatsinas, Chu, Ilyas, and R\'{e}}{Rekatsinas
  et~al\mbox{.}}{2017}]%
        {holoclean}
\bibfield{author}{\bibinfo{person}{Theodoros Rekatsinas}, \bibinfo{person}{Xu
  Chu}, \bibinfo{person}{Ihab~F. Ilyas}, {and} \bibinfo{person}{Christopher
  R\'{e}}.} \bibinfo{year}{2017}\natexlab{}.
\newblock \showarticletitle{HoloClean: Holistic Data Repairs with Probabilistic
  Inference}.
\newblock \bibinfo{journal}{\emph{Proc. VLDB Endow.}} \bibinfo{volume}{10},
  \bibinfo{number}{11} (\bibinfo{date}{aug} \bibinfo{year}{2017}),
  \bibinfo{pages}{1190–1201}.
\newblock
\showISSN{2150-8097}
\urldef\tempurl%
\url{https://doi.org/10.14778/3137628.3137631}
\showDOI{\tempurl}


\bibitem[\protect\citeauthoryear{Ristad and Yianilos}{Ristad and
  Yianilos}{1996}]%
        {levenshtein}
\bibfield{author}{\bibinfo{person}{Eric~Sven Ristad} {and}
  \bibinfo{person}{Peter~N. Yianilos}.} \bibinfo{year}{1996}\natexlab{}.
\newblock \showarticletitle{Learning String-Edit Distance}.
\newblock \bibinfo{journal}{\emph{IEEE Trans. Pattern Anal. Mach. Intell.}}
  \bibinfo{volume}{20} (\bibinfo{year}{1996}), \bibinfo{pages}{522--532}.
\newblock


\bibitem[\protect\citeauthoryear{Singh, Cambronero, Gulwani, Le, Negreanu,
  Raza, and Verbruggen}{Singh et~al\mbox{.}}{2022}]%
        {cornet}
\bibfield{author}{\bibinfo{person}{Mukul Singh}, \bibinfo{person}{Jos{\'e}
  Cambronero}, \bibinfo{person}{Sumit Gulwani}, \bibinfo{person}{Vu Le},
  \bibinfo{person}{Carina Negreanu}, \bibinfo{person}{Mohammad Raza}, {and}
  \bibinfo{person}{Gust Verbruggen}.} \bibinfo{year}{2022}\natexlab{}.
\newblock \showarticletitle{CORNET: A neurosymbolic approach to learning
  conditional table formatting rules by example}.
\newblock \bibinfo{journal}{\emph{arXiv preprint arXiv:2208.06032}}
  (\bibinfo{year}{2022}).
\newblock


\bibitem[\protect\citeauthoryear{Sipser}{Sipser}{1996}]%
        {sipser1996introduction}
\bibfield{author}{\bibinfo{person}{Michael Sipser}.}
  \bibinfo{year}{1996}\natexlab{}.
\newblock \showarticletitle{Introduction to the Theory of Computation}.
\newblock \bibinfo{journal}{\emph{ACM Sigact News}} \bibinfo{volume}{27},
  \bibinfo{number}{1} (\bibinfo{year}{1996}), \bibinfo{pages}{27--29}.
\newblock


\bibitem[\protect\citeauthoryear{University}{University}{2016}]%
        {stanford-edit}
\bibfield{author}{\bibinfo{person}{Stanford University}.}
  \bibinfo{year}{2016}\natexlab{}.
\newblock \bibinfo{title}{CS124 Lecture Notes}.
\newblock
\newblock
\urldef\tempurl%
\url{https://web.stanford.edu/class/cs124/lec/med.pdf}
\showURL{%
\tempurl}


\bibitem[\protect\citeauthoryear{Vaswani, Shazeer, Parmar, Uszkoreit, Jones,
  Gomez, Kaiser, and Polosukhin}{Vaswani et~al\mbox{.}}{2017}]%
        {transformer}
\bibfield{author}{\bibinfo{person}{Ashish Vaswani}, \bibinfo{person}{Noam
  Shazeer}, \bibinfo{person}{Niki Parmar}, \bibinfo{person}{Jakob Uszkoreit},
  \bibinfo{person}{Llion Jones}, \bibinfo{person}{Aidan~N Gomez},
  \bibinfo{person}{\L~ukasz Kaiser}, {and} \bibinfo{person}{Illia Polosukhin}.}
  \bibinfo{year}{2017}\natexlab{}.
\newblock \showarticletitle{Attention is All you Need}. In
  \bibinfo{booktitle}{\emph{Advances in Neural Information Processing
  Systems}}, \bibfield{editor}{\bibinfo{person}{I.~Guyon},
  \bibinfo{person}{U.~Von Luxburg}, \bibinfo{person}{S.~Bengio},
  \bibinfo{person}{H.~Wallach}, \bibinfo{person}{R.~Fergus},
  \bibinfo{person}{S.~Vishwanathan}, {and} \bibinfo{person}{R.~Garnett}}
  (Eds.), Vol.~\bibinfo{volume}{30}. \bibinfo{publisher}{Curran Associates,
  Inc.}
\newblock
\urldef\tempurl%
\url{https://proceedings.neurips.cc/paper/2017/file/3f5ee243547dee91fbd053c1c4a845aa-Paper.pdf}
\showURL{%
\tempurl}


\bibitem[\protect\citeauthoryear{Verbruggen, Le, and Gulwani}{Verbruggen
  et~al\mbox{.}}{2021}]%
        {verbruggen2021semantic}
\bibfield{author}{\bibinfo{person}{Gust Verbruggen}, \bibinfo{person}{Vu Le},
  {and} \bibinfo{person}{Sumit Gulwani}.} \bibinfo{year}{2021}\natexlab{}.
\newblock \showarticletitle{Semantic programming by example with pre-trained
  models}.
\newblock \bibinfo{journal}{\emph{Proceedings of the ACM on Programming
  Languages}} \bibinfo{volume}{5}, \bibinfo{number}{OOPSLA}
  (\bibinfo{year}{2021}), \bibinfo{pages}{1--25}.
\newblock


\bibitem[\protect\citeauthoryear{Wang and He}{Wang and He}{2019}]%
        {unidetect}
\bibfield{author}{\bibinfo{person}{Pei Wang} {and} \bibinfo{person}{Yeye He}.}
  \bibinfo{year}{2019}\natexlab{}.
\newblock \showarticletitle{Uni-detect: A unified approach to automated error
  detection in tables}. In \bibinfo{booktitle}{\emph{Proceedings of the 2019
  International Conference on Management of Data}}. \bibinfo{pages}{811--828}.
\newblock


\bibitem[\protect\citeauthoryear{Wang, Wang, Joty, and Hoi}{Wang
  et~al\mbox{.}}{2021}]%
        {codet5}
\bibfield{author}{\bibinfo{person}{Yue Wang}, \bibinfo{person}{Weishi Wang},
  \bibinfo{person}{Shafiq Joty}, {and} \bibinfo{person}{Steven~C.H. Hoi}.}
  \bibinfo{year}{2021}\natexlab{}.
\newblock \showarticletitle{{C}ode{T}5: Identifier-aware Unified Pre-trained
  Encoder-Decoder Models for Code Understanding and Generation}. In
  \bibinfo{booktitle}{\emph{Proceedings of the 2021 Conference on Empirical
  Methods in Natural Language Processing}}. \bibinfo{publisher}{Association for
  Computational Linguistics}, \bibinfo{address}{Online and Punta Cana,
  Dominican Republic}, \bibinfo{pages}{8696--8708}.
\newblock
\urldef\tempurl%
\url{https://doi.org/10.18653/v1/2021.emnlp-main.685}
\showDOI{\tempurl}


\end{thebibliography}

\end{document}